\documentclass[reprint,
preprintnumbers,
nobibnotes,
superscriptaddress,
 amsmath,amssymb,
 aps,
prd,
floatfix,
]{revtex4-2}

\usepackage{subcaption, caption, graphicx, braket, orcidlink, amssymb, amsmath, dsfont, adjustbox, multirow, lineno, nccmath, scalerel, xcolor, mathtools, booktabs, ulem, bm, hyperref, ragged2e}

\usepackage[linesnumbered,ruled,vlined]{algorithm2e}

\usepackage{dcolumn}
\usepackage{bm}
\usepackage{tikz}
\usetikzlibrary{quantikz2}
\usepackage{cleveref}

\usepackage{physics}

\hypersetup{
    unicode=true,
    pdftoolbar=true,
    pdfmenubar=true,
    pdffitwindow=false,
    pdfstartview={fitH},
    pdftitle={TIMES},
    pdfauthor={J.~Stadelmann, J.~\"{U}belher, M.~Ram\^{o}a , B.~Sambasivam, E.~Barnes, S.E.~Economou},
    pdfsubject={ADAPT-ordering},
    pdfcreator={J.~Stadelmann, J.~\"{U}belher, M.~Ram\^{o}a , B.~Sambasivam, E.~Barnes, S.E.~Economou},
    pdfproducer={},
    pdfkeywords={},
    pdfnewwindow=true,
    colorlinks=true,
    linkcolor=blue,
    citecolor=magenta,
    filecolor=magenta,
    urlcolor=blue
}

\begin{document}


\title{Strategies for Overcoming Gradient Troughs in the ADAPT-VQE Algorithm}

\author{Jonas Stadelmann\orcidlink{0009-0009-4414-1933}}
\email{stadelmannjonas9@gmail.com}
\affiliation{Höhere Technische Lehranstalt (HTL) Bregenz, Vorarlberg, Austria}

\author{Julian \"{U}belher\orcidlink{0009-0002-5848-6181}}
\email{julian.uebelher@gmail.com}
\affiliation{Höhere Technische Lehranstalt (HTL) Bregenz, Vorarlberg, Austria}

\author{Mafalda Ram\^{o}a\orcidlink{0000-0003-0218-7801}}
\email{mafalda@vt.edu}
\affiliation{Department of 
Physics, Virginia Tech, Blacksburg, VA 24061, USA}
\affiliation{Virginia Tech Center for Quantum Information Science and Engineering, Blacksburg, VA 24060, USA}
\affiliation{International Iberian Nanotechnology Laboratory (INL), Portugal}
\affiliation{High-Assurance Software Laboratory (HASLab), Portugal}
\affiliation{Department of Computer Science, University of Minho, Portugal}

\author{Bharath Sambasivam\orcidlink{0000-0002-5765-9469}}
\affiliation{Department of Physics, Virginia Tech, Blacksburg, VA 24061, USA}
\affiliation{Virginia Tech Center for Quantum Information Science and Engineering, Blacksburg, VA 24060, USA}

\author{Edwin Barnes\orcidlink{0000-0002-0982-9339}}
\affiliation{Department of Physics, Virginia Tech, Blacksburg, VA 24061, USA}
\affiliation{Virginia Tech Center for Quantum Information Science and Engineering, Blacksburg, VA 24060, USA}

\author{Sophia E. Economou\orcidlink{0000-0002-1939-5589}}
\email{economou@vt.edu}
\affiliation{Department of Physics, Virginia Tech, Blacksburg, VA 24061, USA}
\affiliation{Virginia Tech Center for Quantum Information Science and Engineering, Blacksburg, VA 24060, USA}


\date{\today}

\begin{abstract}
The adaptive derivative-assembled problem-tailored variational quantum eigensolver (ADAPT-VQE) provides a promising approach for simulating highly correlated quantum systems on quantum devices, as it strikes a balance between hardware efficiency, trainability, and accuracy. Although ADAPT-VQE avoids many of the shortcomings of other VQEs, it is sometimes hindered by a phenomenon known as gradient troughs. This refers to a non-monotonic convergence of the gradients, which may become very small even though the minimum energy has not been reached. This results in difficulties finding the right operators to add to the ansatz, due to the limited number of shots and statistical uncertainties, leading to stagnation in the circuit structure optimization. In this paper, we propose ways to detect and mitigate this phenomenon. Leveraging the non-commutative algebra of the ansatz, we develop heuristics for determining where to insert new operators into the circuit. We find that gradient troughs are more likely to arise when the same locations are used repeatedly for new operator insertions. Our novel protocols, which add new operators in different ansatz positions, allow us to escape gradient troughs and thereby lower the measurement cost of the algorithm. This approach achieves an effective balance between cost and efficiency, leading to faster convergence without compromising the low circuit depth and gate count of ADAPT-VQE.
\end{abstract}

\maketitle


\section{Introduction}\label{sec:Intro}

Calculating the ground state of a quantum many-body system is one of the most fundamental problems in quantum chemistry and physics. After the first calculation of the hydrogen atom orbitals in 1926 by Schrödinger~\cite{Schrodinger1926}, with the rise of classical computing hardware and new calculation schemes, algorithms for calculating the ground state of more complex atoms and molecules were proposed. Classical computational approaches to the problem include mean-field Hartree-Fock theory, density functional theory (DFT), and many more.

Modern quantum chemistry relies on highly optimized algorithms such as DFT~\cite{PhysRev.140.A1133, PhysRev.136.B864}, which maps the many-particle problem to self-consistent single particle equations. Although DFT is helpful to investigate weakly correlated systems, it becomes inaccurate for strongly correlated ones. Other post-Hartree-Fock methods, such as Coupled Cluster (CC) theory~\cite{Cizek1966} (which approximates the ground state by applying the exponential cluster operator, comprised of excitation operators, to the mean-field state), and Full Configuration Interaction (FCI)~\cite{KNOWLES1984315} (which variationally mixes excited and ground mean-field determinants), allow for arbitrary accuracy, but are intractable beyond very small systems.

The idea of using quantum computers for simulating quantum problems was introduced by Feynman in 1982~\cite{Feynman1982}. One of the first and most popular algorithms for this purpose is quantum phase estimation (QPE)~\cite{Whitfield10032011,kitaev1995quantummeasurementsabelianstabilizer}, which for some problems has an exponential advantage with respect to classical counterparts~\cite{PhysRevLett.79.2586, PhysRevLett.83.5162}. However, QPE requires ancillary qubits and the execution of deep circuits that exceed the coherence limits of today's quantum computers, and is therefore only suited for the fault-tolerant quantum computing era~\cite{Preskill2018quantumcomputingin}. Furthermore, QPE requires the initial state to overlap significantly with the target ground state, which necessitates the development of computationally efficient approaches for determining these good initial states.

The variational quantum eigensolver (VQE) was introduced to address these problems~\cite{PeruzzoNature2014, McClean_2016}. Based on the variational principle, VQE substantially reduces circuit depth while allowing for arbitrarily accurate calculations (provided the ansatz is trainable and contains the ground state), and is thus a promising approach for noisy intermediate-scale quantum hardware~\cite{Peruzzo2014, PhysRevX.8.011021, Kandala2017, PhysRevA.95.020501, PhysRevX.8.031022}. VQE only uses the quantum processor for applying gates to a given starting wavefunction and for evaluating the energy and its gradients (if needed) in the resulting state. The variational parameters associated with the gates are tuned by a classical processor to minimize the energy, with the purpose of finding the ground state of the target system.

Despite these advantages, VQEs still face numerous challenges. The accuracy of VQE is highly dependent on the wavefunction ansatz---finding the ground state is only possible if the ansatz is expressive enough to represent a subspace that contains this state~\cite{Gard2020}. For the electronic structure problem, VQEs based on unitary coupled cluster (UCC) theory are a popular option~\cite{RevModPhys.79.291, Yung2014}. However, these typically use unnecessarily deep quantum circuits with many optimization parameters. Possible solutions for reducing the complexity are discussed in Refs.~\cite{doi:10.1021/acs.jctc.8b00932, ryabinkin2019iterativequbitcoupledcluster, doi:10.1021/acs.jctc.8b01004, Huggins_2020, dallairedemers2018lowdepthcircuitansatzpreparing, Romero_2019, Matsuzawa_2020}. Another issue is that Trotterization makes the performance of the ansatz susceptible to changes in operator ordering, with the resulting error fluctuations being relevant at a chemical scale~\cite{doi:10.1021/acs.jctc.9b01083,Evangelista_2019}.

To address these issues, Ref.~\cite{GrimsleyNatureComm2019} introduced ADAPT-VQE, an algorithm that dynamically builds an ansatz that is tailored to the particular system Hamiltonian under study. Starting from a chosen initial approximate wavefunction (i.e., Hartree-Fock), ADAPT-VQE iteratively selects operators with the highest local energy gradient magnitude to sequentially rotate the state towards an energy eigenstate. This gradient criterion leads to the selection of operators that have the potential to significantly lower the energy of the state. In each iteration, the selected unitary is appended to the ansatz with a variational parameter that is optimized in every iteration. Through ADAPT-VQE's problem-tailored and iterative approach, the expressivity of the ansatz is kept minimal while efficiently approaching the ground state.

ADAPT-VQE solves many optimization challenges observed within some other VQEs, such as barren plateaus and local traps~\cite{Grimsley_2023}, while also producing shallower circuits for the same, if not better, accuracy. However, ADAPT-VQE faces some hurdles that prevent implementation on current quantum hardware: The number of measurements required to execute the algorithm is very high, and noise levels resulting from measurements or imperfect gates necessitate the use of error mitigation techniques that lead to additional measurement overheads, hindering near-term implementation for large many-body systems. Many improvements have been proposed in the literature to address these shortcomings. Operator pools based on coupled exchange operators (CEO)~\cite{Ramôa2025}, qubit excitations (QE)~\cite{Yordanov2021}, and individual Pauli strings~\cite{Tang_2021} have been shown to reduce gate counts with respect to the original proposal. Further, gradient measurement protocols~\cite{anastasiouHowReallyMeasure2023b} and optimization improvements~\cite{ramoaReducingMeasurementCosts2024a} have significantly lowered measurement costs. This has resulted in improvements of multiple orders of magnitude with respect to the original proposal~\cite{GrimsleyNatureComm2019}, underlining that through further research, ADAPT-VQE can be improved, potentially to the point where near-term experimental demonstrations for non-trivial problems become feasible.

A yet-unresolved potential roadblock in the implementation of ADAPT-VQE is the presence of gradient troughs~\cite{GrimsleyNPJQI2022}. When a gradient trough occurs, the norm of the gradient falls sharply even though the solution has not been reached. This may result in false convergence or in the incorrect operator being selected due to the difficulty in resolving such small gradients. High computational resources (measurements) are required to select the highest-gradient operators during a gradient trough, and even if the correct operators are selected, the energy will typically not reduce significantly during the gradient trough, leading to an increased ansatz size on top of the measurement overhead. 

One possible approach to remove the unnecessary circuit depth accrued throughout the gradient troughs is to remove operators. This has been attempted in Refs.~\cite{ramoaAnsatzeNoisyVariational2022,vaquerosabater2025prunedadaptvqecompactingmolecularansatze}. Although these heuristic approaches may help compactify the ansatz in certain cases (such as weakly correlated systems or large basis sets), they only apply \textit{after} a gradient trough, and thereby do not reduce the number of iterations required to escape it, nor do they alleviate the measurement costs of choosing good operators along it. Hence, such strategies do not address the main problems stemming from a gradient trough.

In this work, we present methods to diagnose gradient troughs \textit{during} the execution of ADAPT-VQE, allowing them to be properly addressed. We further propose enhanced protocols that are able to escape the gradient troughs. The non-commutativity of the ansatz operators leads to the key insight that inserting the same operator at different points in the ansatz produces distinct gradients and impacts on the energy. To our knowledge, our protocols are the first to use heuristics to select the position in the ansatz to add operators as opposed to always adding them at the end of the ansatz. This introduces new variational paths, leading to a non-vanishing gradient signal throughout the optimization process. This enables ADAPT-VQE to continue to converge even when gradient troughs emerge and the canonical ADAPT-VQE protocol gets trapped.

The rest of this paper is structured as follows. In Sec.~\ref{sec:Background}, we briefly discuss the VQE, ADAPT-VQE and gradient troughs to a more technical extent. Section~\ref{sec:Results} presents the results divided in four parts. In Sec.~\ref{sec:Results_ID_GTs}, we present new techniques to detect gradient troughs during the execution of ADAPT-VQE. We follow with Sec.~\ref{sec:Results_new_protocols}, where we briefly outline our approach and propose protocols aimed at solving the problem of gradient troughs. Furthermore, we compare the cost of each calculation scheme with the original algorithm in Sec.~\ref{sec:Measurement_costs}. In Sec.~\ref{sec:Nummerical_Results}, we numerically investigate the effect of our proposed protocols on gradient troughs. We conclude in Sec.~\ref{sec:Conclusions} with a discussion of the implications of our new protocols and how they might be combined with other methods to obtain an efficient and robust version of ADAPT-VQE. Finally, we include additional details in the appendix. In App.~\ref{app:AppA}, we provide a thorough derivation of the gradient formula for adding operators into general ansatz positions. In App.~\ref{app:AppB}, we include pseudocodes describing the protocols introduced in the main text. App.~\ref{app:AppC} provides a detailed exposition on the correlation between gradient troughs and operator gradients across the ansatz.


\section{Background}\label{sec:Background}

\subsection{Variational Quantum Eigensolver}\label{subsec:VQE_Background}

VQE is a ground state preparation algorithm based on the variational principle of quantum mechanics. In this algorithm, the expectation values of the Hamiltonian operator are determined by a quantum processor, while the circuit parameters are tuned by a classical processor to minimize the Rayleigh–Ritz quotient:

\begin{align}
E(\vec{\theta})=\frac{\bra{\psi (\vec\theta)} \hat{H} \ket{\psi (\vec\theta)}
}{\bra{\psi (\vec\theta)} \ket{\psi (\vec\theta)}}.   \label{eq:Rayleigh–Ritz-quotient}
\end{align}

By the variational principle, the state that minimizes this quantity is the one of minimal energy---the ground state of $H$. Hence, provided the energy optimization converges to the solution and the ansatz is expressive enough to represent the ground state, the ground state will be reached. Upon termination of the algorithm, the optimized parameters are stored as a reproducible description of the ground state approximation. 

In any variational algorithm, the ansatz is a fundamental component: The structure of the variational wavefunction dictates the quality of the solution within reach. In the context of VQE, one of the first and most popular proposals was the UCCSD (unitary coupled cluster single and doubles) ansatz \cite{PeruzzoNature2014, PhysRevA.95.020501, Romero_2019, doi:10.1021/acs.jctc.9b01083, xia2020qubitcoupledclustersingles, doi:10.1021/acs.chemrev.8b00803}. This ansatz uses physically motivated operators, inspired by classical variational methods for quantum chemistry. The most common VQEs, including UCCSD-VQE, are static: they implement a quantum circuit with a fixed gate structure, and only the gate parameters are modified during the execution. This results in a significant number of operators that do not contribute to reaching the target wavefunction, creating an unnecessarily deep quantum circuit and imposing a high optimization cost. In addition to this, a subset of static ans\"{a}tze suffer from barren plateaus, a problem characterized by the exponential decay of gradients and the consequent impracticality of optimization for substantial system sizes \cite{McClean2018, Cerezo2021BarrenPlateaus}. Finally, in the context of the prevalent totterized implementations of UCCSD, this ansatz has been shown to not be chemically well-defined \cite{doi:10.1021/acs.jctc.9b01083}.

\subsection{ADAPT-VQE}\label{subsec:ADAPT-VQE_Background}

Ref.~\cite{GrimsleyNatureComm2019} introduced ADAPT-VQE to address the limitations of fixed ans\"{a}tze, such as UCCSD, by building the ansatz dynamically. Provided an initial approximate wavefunction (such as the Hartree-Fock ground state in chemistry), ADAPT-VQE selects operators one by one based on local energy gradients. Specifically, in each iteration, the operator corresponding to the largest absolute energy gradient with respect to the corresponding variational parameter set to zero  (Eq.~\eqref{eq:ADAPT_derivative_appending}) is selected. These operators are chosen from a predefined pool consisting of problem- and reference-state-tailored excitation operators. This method ensures that only operators with the potential to significantly decrease the energy are included in the ansatz. After the chosen operator is added, a full optimization of all the parameters in the ansatz is performed. The operator selection procedure is repeated after each optimization. While all parameters are re-optimized in each iteration, the values of the previous parameters serve as a warm start for the optimization, benefiting trainability and reducing optimization cost. New parameters are initialized to zero.

The parameterized state after $n$ ADAPT-VQE iterations is defined by:
\begin{align}
\ket{\psi_{n}} =& e^{{\theta_n}{A_n}} e^{{\theta_{n-1}}{A_{n-1}}} \text{...}  e^{{\theta_{2}}{A_{2}}} e^{{\theta_1}{A_1}}  \ket{\psi_{0}}  \label{eq:ADAPT_append},
\end{align}
where the $A_j$ represent anti-hermitian operators selected from an operator pool. The original ADAPT-VQE algorithm consistently \textit{appends} operators to the ansatz, i.e., adds them to the end, as represented by the $n$th operator being to the left in Eq.~\eqref{eq:ADAPT_append}. However, it is also conceivable to \textit{prepend} them, i.e., add new operators to the beginning of the ansatz (placing them to act directly on the reference state). In between these extremes, there are many other options ($n$ possible positions for an $n$-operator ansatz). While this has not been previously explored in the ADAPT-VQE literature, we will generalize the ADAPT-VQE operator addition to include these possibilities.

The operator gradients required to select a new ansatz element for appending are given by the following expectation values, which are evaluated on quantum hardware:
\begin{align}
\left.\frac{\partial E}{\partial\theta_{j}}\right|_{\theta_{j} = 0}
= \bra{\psi_{n}} [\hat{H}, A_{j}] \ket{\psi_{n}}. \label{eq:ADAPT_derivative_appending}
\end{align}
This expectation value must be measured for every operator $A_j$ in the pool.

The ADAPT-VQE algorithm terminates when a certain convergence criterion is met, which typically corresponds to the norm of the vector containing all pool operator gradients falling below a predefined threshold.

Similar to other VQEs, ADAPT-VQE evaluates the energy expectation values and their gradients on quantum hardware, leaving the variational optimization up to a classical processor that receives feedback from the quantum computer. However, in ADAPT-VQE, the ansatz structure itself is dictated by heuristics based on measurements performed during the execution, leading to a system-tailored circuit. This circuit depends on the operator pool, which must be defined carefully to be able to represent the ground state and is typically motivated by the problem at hand~\cite{Ramôa2025,Yordanov2021,Tang_2021,GrimsleyNatureComm2019}.

\subsection{Gradient Troughs}\label{subsec:Gradient_Troughs}

\begin{figure}[!htbp]
    \centering
    \includegraphics[width=0.45\textwidth]{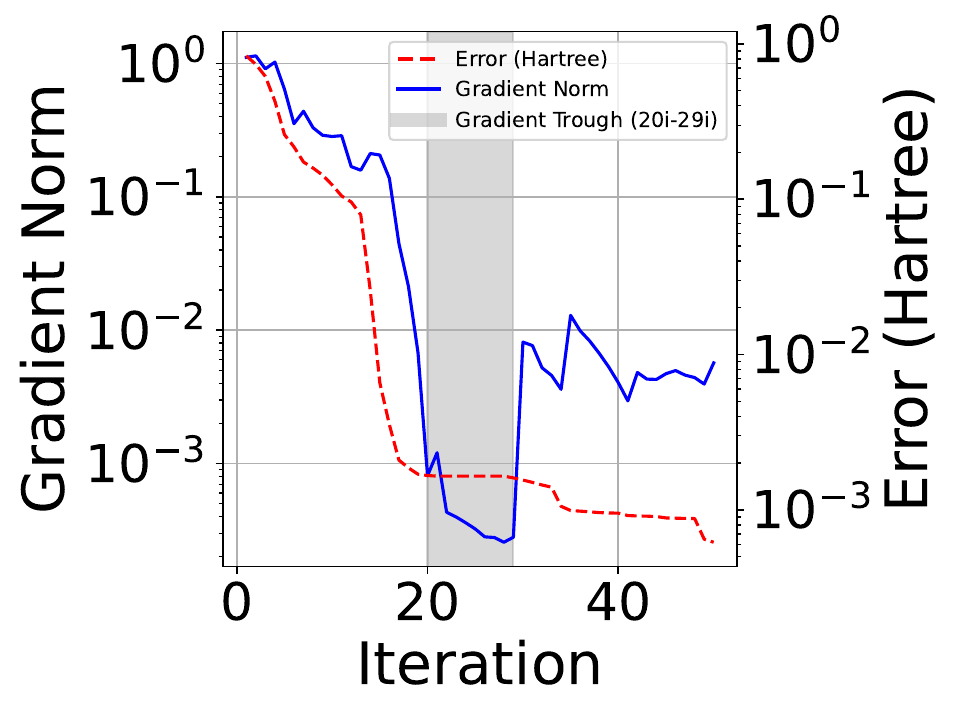}
    \caption{\justifying Noise-free simulation of ADAPT-VQE for a linear H$_6$ molecule with $4\text{ \AA}$ interatomic distance until iteration 50. Between iterations $20$ and $29$, a significant gradient trough is evident, accompanied by a substantial decrease in gradient norm and a convergence plateau in energy, as is typical for them. The error is defined with respect to the FCI energy.}
    \label{fig:gradient_trough_normal_calculation}
\end{figure}

Although ADAPT-VQE resolves optimization issues such as barren plateaus and susceptibility to local traps, it encounters a related problem known as ``gradient troughs''. A gradient trough refers to a regime during the optimization in which the gradient norms decrease rapidly even though the ground state has not yet been reached, only to increase again after some iterations. In classical simulations with effectively unlimited precision in the energy evaluation, this results in the stalling of the energy and with it error, as shown in Fig.~\ref {fig:gradient_trough_normal_calculation}, which in practice amounts to the allocation of both quantum as well as classical computational resources without a measurable effect on the reduction of the energy. In a realistic setting, the low gradient norms within the troughs might lead to premature and wrongful convergence if the value drops below the threshold that defines termination of the algorithm. Furthermore, gradient troughs increase measurement costs: When gradients become very small, they must be measured to high precision to ensure that the operator with the highest gradient magnitude is added to the ansatz within the ADAPT-VQE selection step.

Because both relate to low gradients, there is a common misconception that gradient troughs in ADAPT-VQE are conceptually equivalent to barren plateaus in VQE. This is not the case. Barren plateaus are characterized by gradients that vanish exponentially on the system size on average for random points in parameter space \cite{McClean2018}. They hinder the ability to scale up the optimization to larger system sizes and threaten the practical viability of VQE, as they lead to exponential optimization costs. Barren plateaus stem from random ansatz structures and a lack of good heuristics for parameter initialization. 

In ADAPT-VQE, we have good initialization techniques (based on the Hartree-Fock solution and parameter recycling) and the ansatz is not random; rather, it's informed both by the general problem (through fermion-tailored pools) and the particular system at hand (through the Hamiltonian-driven selection criterion). Furthermore, the algorithm is designed to consistently add operators with high gradient magnitudes. We have theoretical and numerical reasons to believe that the VQE subroutine in ADAPT-VQE does not suffer from barren plateaus \cite{GrimsleyNPJQI2022}. 

One might ask if gradient troughs are not a manifestation of the same problem. It is conceivable that instead of observing exponential vanishing gradients during the optimization, we would observe them during the operator selection step. However, chemistry-based considerations related to size-consistency suggest that gradient troughs do not deepen with system size \cite{GrimsleyNPJQI2022}. Gradient troughs are observed numerically, not proved to exist analytically; they only arise for a subset of molecular systems, and are related to low-lying energy eigenstates rather than particular characteristics or shortcomings of the ansatz and parameter initialization. While a rigorous study of the scaling of the gradient trough phenomenon is relevant, it is outside the scope of this work, where we focus on heuristic methods for mitigating gradient troughs as observed in numerical simulations.


\section{Results}\label{sec:Results}

During the execution of ADAPT-VQE for strongly correlated systems, we often observe the recurrent appearance of gradient troughs. Within these regions, the energy stalls and gradients become very small. As discussed in Sec.~\ref{subsec:Gradient_Troughs}, this may cause the algorithm to converge prematurely and/or require very high measurement precision in the operator selection process (increasing costs). This poses a significant problem for the practical deployment of ADAPT-VQE, as we have no systematic way of distinguishing between gradient troughs and convergence, given realistic constraints on measurement costs. 

The standard ADAPT-VQE algorithm appends operators to the end of the ansatz. A gradient trough is a statement on the smallness of the operator pool gradients in this ``appending'' position. We could also look at the operator pool gradients associated with inserting operators at other positions in the ansatz. Based on numerical simulations, we will show that, during a gradient trough, gradients tend to be smaller in positions close to the end of the ansatz (i.e., when appending). In other positions, the gradients can be higher, making them easier to measure and signaling that convergence has not been reached. In this section, we will use this observation to develop several heuristic protocols to detect and mitigate gradient troughs in ADAPT-VQE.

Throughout this paper, our numerical calculations focus exclusively on the linear H$_6$ molecule with an interatomic distance of $4\text{ \AA}$. This is a typical toy problem that serves as a proxy for strongly correlated systems and suffers from gradient troughs.  The basis set is STO-3G, such that the test system has 12 qubits. In all cases, we take the error with respect to the FCI energy, which is exact within the chosen basis set. Regarding the operator pool, we used the problem-specific qubit excitation (QE) pool in all cases~\cite{Yordanov2021}. We allow the algorithm to run for as many iterations as allows us to observe the relevant features of a gradient trough.

In what follows, we introduce and compare four distinct protocols for inserting operators. The protocols differ on how the operator and its position within the ansatz are chosen. This will allow us to obtain a more comprehensive picture of gradient trough detection and mitigation. 

\begin{figure*}[htb]
  \centering

  \begin{subfigure}[t]{0.48\textwidth}
    \centering
    \includegraphics[width=\linewidth]{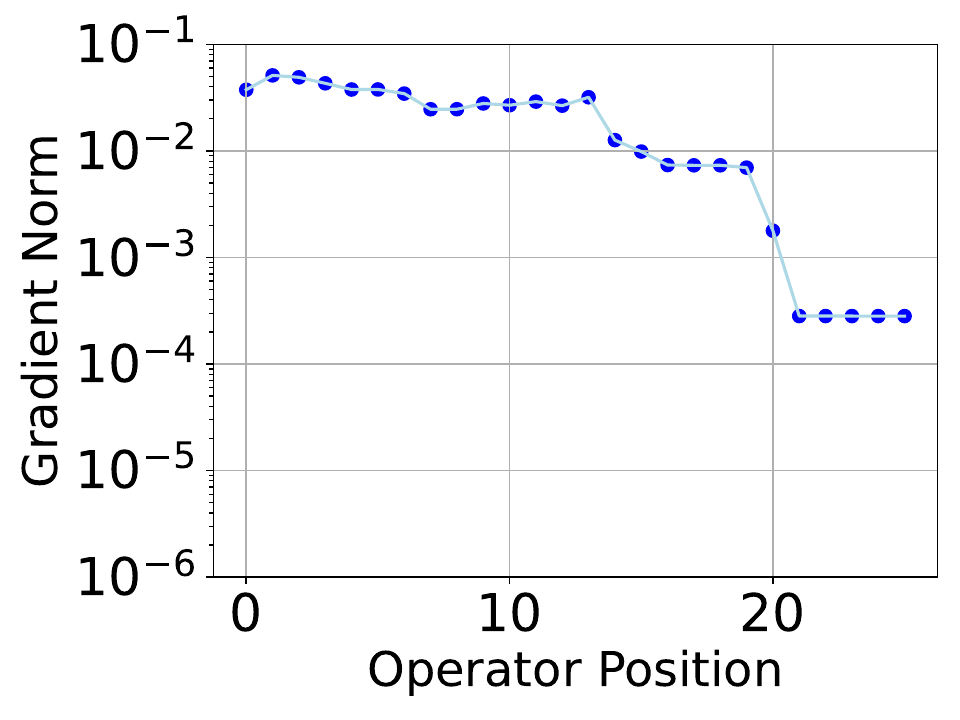}
    \caption{\justifying Gradient norm versus operator position inside a gradient trough (iteration 25).}
    \label{fig:suba}
  \end{subfigure}
  \hfill
  \begin{subfigure}[t]{0.48\textwidth}
    \centering
    \includegraphics[width=\linewidth]{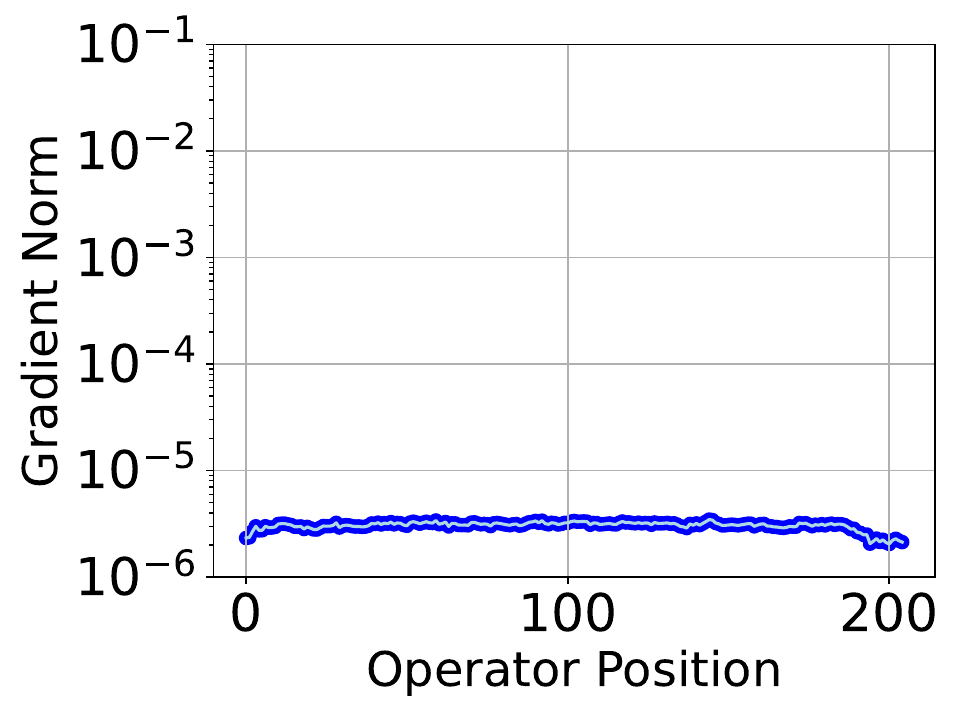}
    \caption{\justifying Gradient versus operator at true convergence (iteration 205) with a convergence threshold of $5*10^{-7}$ for the appending position.}
    \label{fig:subb}
  \end{subfigure}

  \caption{\justifying Gradient norm against operator position for a regular execution of ADAPT-VQE (i.e., appending operators). Panel (a) shows an iteration within a gradient trough, while the right panel considers an iteration at convergence. Within the gradient trough, a clear decrease in the gradient norm from the prepending to the appending position can be observed. For the case at convergence, there is no such consistent decrease visible: Instead, we can observe that the gradient norms are mostly uniform across positions.}
  \label{fig:gradient_norm_over_operator_positions}
\end{figure*}

\subsection{Identification of Gradient Troughs}\label{sec:Results_ID_GTs}

In this section, we propose a new protocol to detect gradient troughs with the primary goal of distinguishing them from true convergence of the algorithm. Our protocol involves the measurement of the gradient norm of the pool operators at various positions $p$ within the ansatz. As discussed earlier, the gradient norm becomes small when $p$ is close to $n+1$ (the end of the ansatz). On the other hand, when $p$ is farther away from the end of the ansatz, the gradient norm can be much larger, having a greater impact on the cost function landscape. This phenomenon can be elucidated by numerically comparing the gradient formulas for the appending position with those for other positions within the ansatz. If the algorithm has indeed truly converged, the gradient norm across all $p$ will be small. 

We show a numerical example of this difference between a gradient trough and true convergence in Fig.~\ref{fig:gradient_norm_over_operator_positions}. In Fig.~\ref{fig:suba}, the total gradient norm of all operators within the pool is plotted against the ansatz positions for an iteration inside a gradient trough. The gradient norm decreases steadily as the operator insertion point moves from the beginning to the end of the ansatz. When we compare this to true convergence shown in Fig.~\ref{fig:subb}, a major difference can be noted: In the converged case, the gradient norms are much more uniform across the positions, not showing the same decrease we observed within the troughs (Fig.~\ref{fig:suba}).

In the context of the appending position, the gradient of a pool operator $A_{\mu}$ is given by Eq.~\eqref{eq:ADAPT_derivative_appending} and corresponds to the expectation value of the commutator [$\hat{H}$, $A_\mu$] in the current state. On the other hand, the gradient of an operator $A_{\mu}$ in a general position $p$ within the ansatz can be defined as:

\begin{align}
\left.\frac{\partial E}{\partial\theta_\mu}\right|_{\theta_\mu=0}
&=
\bra{\psi}
\bigl[\hat{H},\;
V(\theta)
A_\mu
V^\dagger(\theta)
\bigr]
\ket{\psi},
\label{eq:ADAPT-energy_gradient_any_position}
\end{align}

\begin{align}
V(\theta)
&=
\left(\prod_{i=p}^{n} e^{\theta_{p+n-i} A_{p+n-i}}\right)
\label{eq:V_gradient_any_position}
\end{align}
where $p$ must be within the range of position $p=1$ (prepending) up to position $p=n+1$ (appending), and the products are defined such that factors corresponding to smaller values of $i$ are on the left, while those with larger values of $i$ are on the right. The detailed derivation of this formula is provided in Sec.~\ref{app:AppA}.

Therefore, the presence of low gradient norms in the appending position, as observed in gradient troughs, does not inherently imply that the gradient norms are small for all ansatz positions. This is due to the fact that the gradients for non-appending positions involve an additional conjugation of $A_\mu$ by V($\theta$), which depends on subsequent variational layers. A concise numerical investigation of this is furnished in Sec.~\ref{app:AppC}.

\begin{figure}[!htbp]
    \centering
    \includegraphics[width=0.5\textwidth]{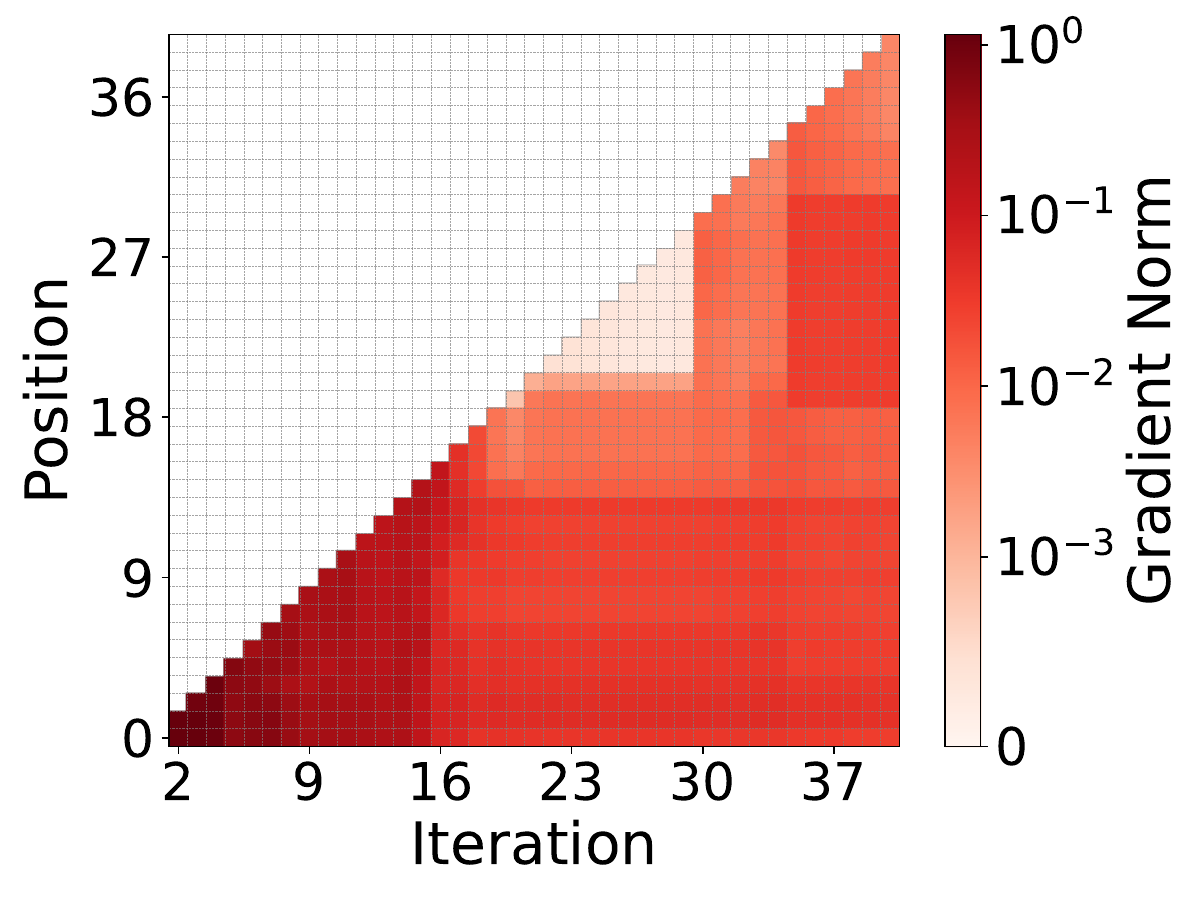}
    \caption{\justifying Gradient norm across ansatz positions for ADAPT-VQE iterations 1--40. The gradient trough is visible across iterations 20--29 as a decrease in gradient norm as the position moves towards the end of the ansatz (with appending being the extreme case).}
    \label{fig:stair_plot_gradient_trough}
\end{figure}

This property of gradient troughs can be observed numerically in Fig.~\ref{fig:stair_plot_gradient_trough}. The plot clearly shows that before the trough, the gradient norm for adding an operator in a specific position of the ansatz has a similar order of magnitude across all positions. In contrast, as the algorithm enters the gradient trough (iteration 20), the gradient norms in the appending position (top) are significantly lower compared to the gradient norms for the prepending position (bottom). As the algorithm approaches iteration 30 and exits the gradient trough, the gradient norms are again within the same order of magnitude across the ansatz. A parallel can be made with Fig.~\ref{fig:gradient_trough_normal_calculation}, which corresponds to the gradient norm in the appending position alone, as both plots exhibit the same gradient trough.

Figures~\ref{fig:gradient_norm_over_operator_positions} and \ref{fig:stair_plot_gradient_trough} provide us with a crucial tool to distinguish between convergence and a gradient trough. To make sure the algorithm has truly converged to the ground state, we must check that all gradient norms are small and of similar magnitude across ansatz positions. If instead we observe that the gradient norms decrease significantly towards the end of the ansatz, we are inside a gradient trough. Using these results, one may redefine the convergence criterion as having all the gradient norms for all ansatz positions below a specific threshold, beyond which it is not practical to resolve the gradients with sufficient accuracy. This guarantees that the algorithm cannot further decrease the energy, as it has either attained (within the available accuracy) a local minimum or, conceivably, the global minimum. Current convergence criteria are based exclusively on the gradient norm in the last ansatz position (appending) falling below a preset threshold, which risks converging too early due to a gradient trough. However, this approach has the benefit of requiring fewer measurements, as measuring the gradient norm over all ansatz positions incurs a significant measurement overhead. 

To efficiently implement a method that diagnoses convergence and avoids premature termination, a tradeoff between the two convergence criteria may be developed and implemented. This will allow for a realistic incorporation of our findings into the ADAPT-VQE protocol, allowing for the detection of troughs and preventing false convergence while minimizing the measurement overhead. We leave the specific implementation of such a method for future work; however, we outline the basic framework below. 

Before a given convergence criterion (based on the appending gradient norm) is met, the regular ADAPT-VQE protocol is executed. Once the gradient norm falls below the convergence threshold, the norm for a select number of other positions within the ansatz is measured. The choice of positions to test can be random or based on heuristics. From Fig.~\ref{fig:stair_plot_gradient_trough}, we can see that the operator gradients are non-vanishing for any ansatz position that is not within a few positions of the appending position. One possibility is to restrict the test positions by selecting one (or more) at random with a bias towards the beginning of the ansatz (by, e.g., creating a probability distribution over positions that decays with the distance to the beginning of the ansatz). After measuring the gradients in the test positions, we may conclude that a gradient trough is present if the gradient norm is significantly higher in positions other than appending, or (given enough data points) if the measurements show a decline in the gradient norms as the default position for adding new operators (end of ansatz) is approached. If the difference between the norm in the appending position and in other positions is within a preset threshold, we conclude that the algorithm has terminated and the ground state has been found.

\subsection{New Protocols}\label{sec:Results_new_protocols}

While the discussion in Sec.~\ref{sec:Results_ID_GTs} proposed tools to diagnose gradient troughs, it did not provide tools to mitigate their effects on the ADAPT-VQE algorithm. In this section, we introduce a set of new protocols designed to escape gradient troughs. These protocols are aimed at ensuring a consistent optimization process even when a gradient trough occurs within the regular ADAPT-VQE calculation. They make use of the non-commutative properties of the exponential operators, which result in different gradients at each ansatz position for each candidate operator. 

We have observed the variation of the gradients for different ansatz positions in Fig.~\ref{fig:stair_plot_gradient_trough}. The canonical ADAPT-VQE algorithm always appends operators to the ansatz, but as the algorithm enters the gradient trough, the gradient norms for the appending position become smaller, causing the optimization problems highlighted in Sec.~\ref{subsec:Gradient_Troughs}. On the other hand, the gradients for inserting operators in other ansatz positions remain high due to the effects of the ansatz position on the commutator within the gradient formula (Eq.~\eqref{eq:ADAPT-energy_gradient_any_position}). The protocols we propose leverage the possibility of adding new operators into ansatz positions other than appending. The ansatz for one of the new protocols can be expressed as:

\begin{equation}
\begin{aligned}
\ket{\psi_{\text{ansatz}}}
&=
\left(\prod_{i=p}^{n} e^{\theta_{p+n-i} A_{p+n-i}}\right)
e^{\theta_\mu A_\mu} \\[1em]
&\quad
\left(\prod_{i=1}^{p-1} e^{\theta_{p-i} A_{p-i}}\right)
\ket{\psi_0},
\end{aligned}
\label{eq:ADAPT_ansatz_new_protocols}
\end{equation}
where we introduced the new operator $e^{\theta_\mu A_\mu}$ into position $p$ within the ansatz. In the typical ADAPT-VQE algorithm, $p=n+1$ always. The energy expectation value on this parametrized variational wavefunction is given by

\begin{equation}
\begin{aligned}
E
&= \bra{\psi_0}
\left(\prod_{i=1}^{p-1} e^{-\theta_i A_i}\right)
e^{-\theta_\mu A_\mu}
\left(\prod_{i=p}^{n} e^{-\theta_i A_i}\right)
\hat{H} \\[1em]
&\quad
\left(\prod_{i=p}^{n} e^{\theta_{p+n-i} A_{p+n-i}}\right)
e^{\theta_\mu A_\mu}
\left(\prod_{i=1}^{p-1} e^{\theta_{p-i} A_{p-i}}\right)\\[1em]
&\quad
\ket{\psi_0}.
\end{aligned}
\label{eq:ADAPT-energy_expectation_value_mu_in_main_text}
\end{equation}

\begin{figure*}[p] 
    \centering
    \includegraphics[width=\textwidth]{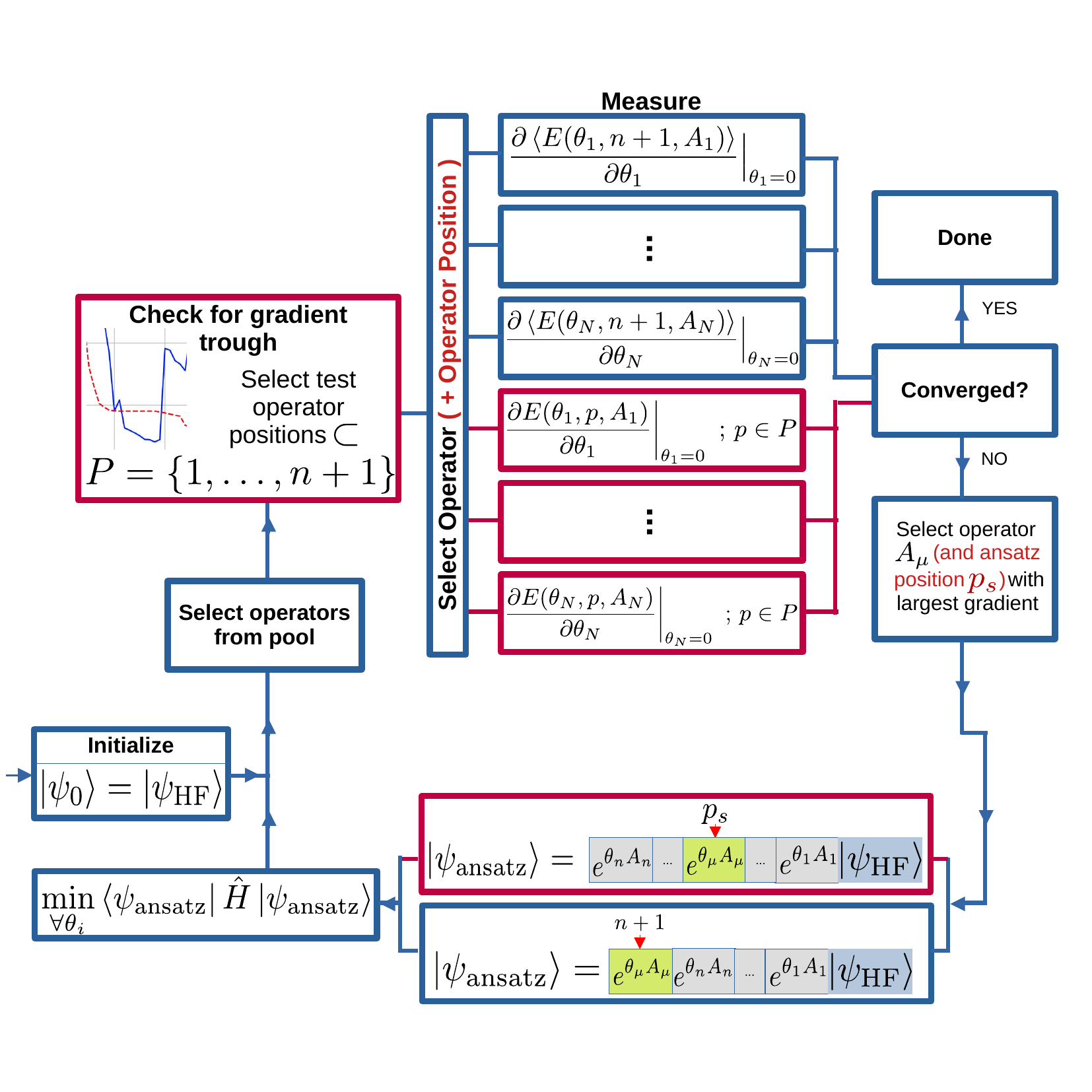}
    \caption{\justifying 
        Workflow of the enhanced ADAPT-VQE algorithms, where red indicates our changes to the protocols and blue is the original ADAPT-VQE algorithm as stated in Sec.~\ref{subsec:ADAPT-VQE_Background}. Unlike the original ADAPT-VQE, our enhanced algorithms check for gradient troughs. If a gradient trough occurs, the protocol measures gradients at positions that are most likely to yield higher gradients than appending. In the next step, the operator and position that yield the largest magnitude gradient are identified, and the operator is inserted accordingly into the ansatz. All three algorithms are based on this principle. For an exact description see Sec.~\ref{sec:Results_new_protocols}. The function $E(\theta_\mu, p_s, A_\mu)$ refers to the energy expectation value of the system as defined in Eq.~\ref{eq:ADAPT-energy_expectation_value_mu}, where in the given case $\theta_\mu$ is the variational parameter for the operator $A_\mu$, $p_s$ refers to the position of the newly added operator $A_\mu$ in the ansatz and $A_\mu$ with $\mu \in [1,N_\text{Pool}]$ is defined to be an operator within a given pool of size $N_\text{Pool}$.
    }
    \label{fig:algorithm_scheme}
\end{figure*}

In this work, we introduce four protocols: random operator random position (RO/RP), random operator optimized position (RO/OP), optimized operator random position (OO/RP) and optimized operator optimized position (OO/OP). The implementation of these protocols only requires minor changes to the original (appending) algorithm. We depict them in Fig.~\ref{fig:algorithm_scheme} and describe each protocol in detail below. The pseudocodes for the protocols are provided in Sec.~\ref{app:AppB}.\\

\begin{itemize}
    \item \textbf{OO/OP (optimized operator optimized position)}: The energy gradient for a fixed number of pool operators (here taken to be 10) is measured for all possible ansatz positions. The 10 operators whose gradients across the ansatz we choose to measure are those with the highest gradient for appending. Out of the 10 operators and all ansatz positions, we select the combination of operator and position associated with the highest gradient.
    
    \item \textbf{OO/RP (optimized operator random position)}: The ansatz position is selected randomly, and the gradients of all pool operators are measured in this position. We select the operator with the largest energy gradient magnitude for this position.
    
    \item \textbf{RO/OP (random operator optimized position)}: The energy gradient for a fixed number of pool operators is measured for all possible ansatz positions. Here we choose this number to be $10$. The $10$ operators whose gradients across the ansatz we choose to measure are those with the highest gradient for appending. The selected position is the one where these gradients are highest. The operator to be added is selected randomly and placed into the identified position. This protocol should be seen as a control for the OO/OP protocol that we use to understand how important it is to select the highest gradient operator, as compared to a random operator, when the position is selected \textit{via} our heuristics. As compared to OO/OP, RO/OP selects the operator from the complete pool, rather than being restricted to the top 10 operators for appending in a specific iteration.
    
    \item \textbf{RO/RP (random operator random position)}: Both the position and the new ansatz element are chosen randomly.
\end{itemize}

All these protocols make use of the fact that adding operators in other positions in the ansatz can lead to a higher impact on the cost function landscape, thereby helping escape or avoid gradient troughs. When we encounter such a problem, the gradients of operators will become low for appending (Eq.~\eqref{eq:ADAPT_derivative_appending}). However, by adding operators in different ansatz positions (Eq.~\eqref{eq:ADAPT-energy_gradient_any_position}), the gradient norm is significantly increased in most cases. Using our protocols to select an operator and its position as a gradient trough occurs, the algorithm is able to escape/avoid it relatively quickly. The apparent price of the application of these algorithms is that, except for the random operator random position (RO/RP) protocol, these algorithms seemingly require a higher computational cost compared to the default ADAPT-VQE protocol. We follow with a comparison of the measurement costs.

\subsection{Measurement costs}\label{sec:Measurement_costs}

In the context of a realistic implementation on a quantum computer, it is essential to incorporate the statistical deviation coming from measuring expectation values with a finite number of shots. To obtain sufficient accuracy, we expect it to be necessary to measure each gradient approximately $\mathcal{O}\left(\frac{1}{\lVert g \rVert_\infty^2}\right)$ times, where $\lVert g \rVert_\infty$ is the infinity norm of the gradient vector, since this is the scaling of the number of shots required to distinguish the highest pool operator gradient from zero. For qubit or fermionic excitations acting on real states (as is our case), it has been shown that the cost of measuring each gradient is roughly equivalent to the cost of measuring the energy (requiring only a constant number of additional gates), just like the case of a single Pauli string~\cite{D0SC06627C}. Thus, we can estimate the number of shots required to measure all pool operator gradients in any given ansatz position, $N_\text{g\_meas}$, as

\begin{align}
N_\text{g\_meas} &= \mathcal{O}\left(\frac{N_\text{H} N_\text{Pool}}{\lVert g \rVert_\infty^2
} \right)
\label{eq:general_operator_measurement_cost}
\end{align}
where $N_\text{H}$ is the number of Pauli strings in the Hamiltonian and $N_\text{Pool}$ is the number of operators in the pool. For molecular Hamiltonians and the pool we are considering, we have that $N_\text{H}=N_\text{Pool}=\mathcal{O}\left(N_\text{orbs}^4\right)$, where $N_\text{orbs}$ is the number of orbitals, leading to a scaling of $\mathcal{O}\left(\frac{N_\text{orbs}^8}{\lVert g \rVert_\infty^2} \right)$ for the number of shots required. 

We may already note that, despite seemingly incurring higher measurement costs, our protocols are actually capable of saving quantum resources on the ADAPT-VQE operator selection step by decreasing the precision requirements via an increase in the gradient norm $\lVert g \rVert_\infty$, which typically drops drastically (thereby requiring higher measurement costs) in a gradient trough. In our simulations, we observed it to be typical for our protocols to increase the gradient norm by two orders of magnitude with respect to the original protocol within a gradient trough, leading to savings on the order of $10^4$ in the measurement costs required to evaluate the gradients in each iteration. This difference will be exacerbated by the parameter optimization, since, during a gradient trough, more measurements will be required to repeatedly resolve the search direction.

While measuring the appending operator gradients might come at a lower cost in our protocols, most of them require measuring gradients in additional ansatz positions, which results in additional quantum computer calls. More precisely, the cost of selecting an operator and position for the four protocols we propose is given by

\begin{equation}
\begin{aligned}
N_\text{m-oo/op} &= N_\text{g\_meas} + 10 N_\text{Positions} N_\text{H}\\
&=\mathcal{O}\left(\frac{N_\text{orbs}^8}{\lVert g \rVert_\infty^2} \right)+\mathcal{O}\left(\frac{N_\text{orbs}^4N_\text{Positions}}{\lVert g \rVert_\infty^2} \right), \\
N_\text{m-oo/rp} &= N_\text{g\_meas}=\mathcal{O}\left(\frac{N_\text{orbs}^8}{\lVert g \rVert_\infty^2} \right)\\
N_\text{m-ro/op} &= N_\text{g\_meas} + 10 N_\text{Positions}  N_\text{H}\\
&=\mathcal{O}\left(\frac{N_\text{orbs}^8}{\lVert g \rVert_\infty^2} \right)+\mathcal{O}\left(\frac{N_\text{orbs}^4N_\text{Positions}}{\lVert g \rVert_\infty^2} \right), \\
N_\text{m-ro/rp} &= 0,
\end{aligned}
\label{eq:protocolspecific_operator_measurement_cost}
\end{equation}
where the subscripts indicate the protocol and $N_\text{Positions}$ corresponds to the number of positions we are testing within the ansatz.

We observe that only the protocols OO/OP and RO/OP incur a measurement overhead with respect to the regular ADAPT-VQE operator selection cost given by Eq.~\eqref{eq:general_operator_measurement_cost}. This cost is dependent on the protocol-specific hyperparameter $N_\text{Positions}$. In our simulations, we chose to test all possible positions within the ansatz. This is typically significantly lower than the number of operators in the pool, meaning that the additional costs are eclipsed by the typical cost of operator selection in ADAPT-VQE, $\mathcal{O}\left(\frac{N_\text{orbs}^8}{\lVert g \rVert_\infty^2} \right)$. Further, the value of $\lVert g \rVert_\infty^2$ will be much higher for our protocols during a gradient trough, leading to an effective overall decrease in measurement costs.

To further lower the cost, we could only measure gradients for a select subset of ansatz positions, thereby decreasing $N_\text{Positions}$. As we discussed in Sec.~\ref{sec:Results_ID_GTs}, one interesting possibility is to select positions randomly with a bias towards the beginning of the ansatz. We leave the exploration of such ideas for future work.

Finally, we note that when we append operators, we can use Eq.~\eqref{eq:ADAPT_derivative_appending} to obtain the gradients, which has the form of an expectation value for all pool operators. This opens the door to strategies that group the observables into commuting sets, which may bring the cost down from $\mathcal{O}\left(\frac{N_\text{orb}^8}{\lVert g \rVert_\infty^2} \right)$ to $\mathcal{O}\left(\frac{N_\text{orb}^5}{\lVert g \rVert_\infty^2} \right)$ for molecular Hamiltonians \cite{anastasiouHowReallyMeasure2023b}. In contrast, adding operators at other ansatz positions necessarily implies using parameter-shift rules, leading us to the scaling of Eq.~\eqref{eq:general_operator_measurement_cost}. However, from our cost estimates in Eq.~\eqref{eq:protocolspecific_operator_measurement_cost} and the discussion above, we can see that the asymptotic costs of our protocols would still match those of the original algorithm, or improve upon them by increasing $\lVert g \rVert_\infty^2$. Further, our protocols and cost analysis are not restricted to molecular Hamiltonians and can be readily applied to any instance of ADAPT-VQE.

\subsection{Impact of the Proposed Protocols}\label{sec:Nummerical_Results}
In order to quantitatively address our proposed protocols, we investigate their impact on the gradient trough we observed in Sec.~\ref{sec:Results_ID_GTs}. First, we will apply the new protocols for a fixed number of iterations, three iterations after the start of the gradient trough. Subsequently, an investigation will be conducted to ascertain whether the algorithm exits the trough or re-enters it.

\begin{figure*}[p] 
\centering

\begin{minipage}{0.45\textwidth}
    \centering
    \includegraphics[width=\linewidth]{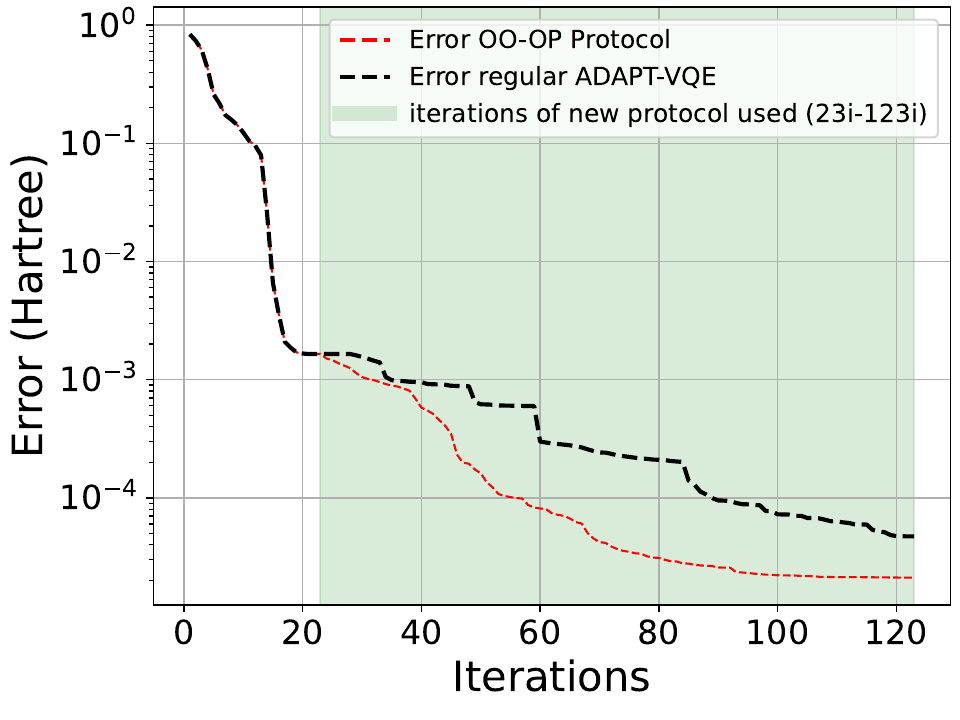}
    \caption*{\justifying (a) OO/OP Protocol (optimized operator, optimized position)}
\end{minipage}
\hfill
\begin{minipage}{0.45\textwidth}
    \centering
    \includegraphics[width=\linewidth]{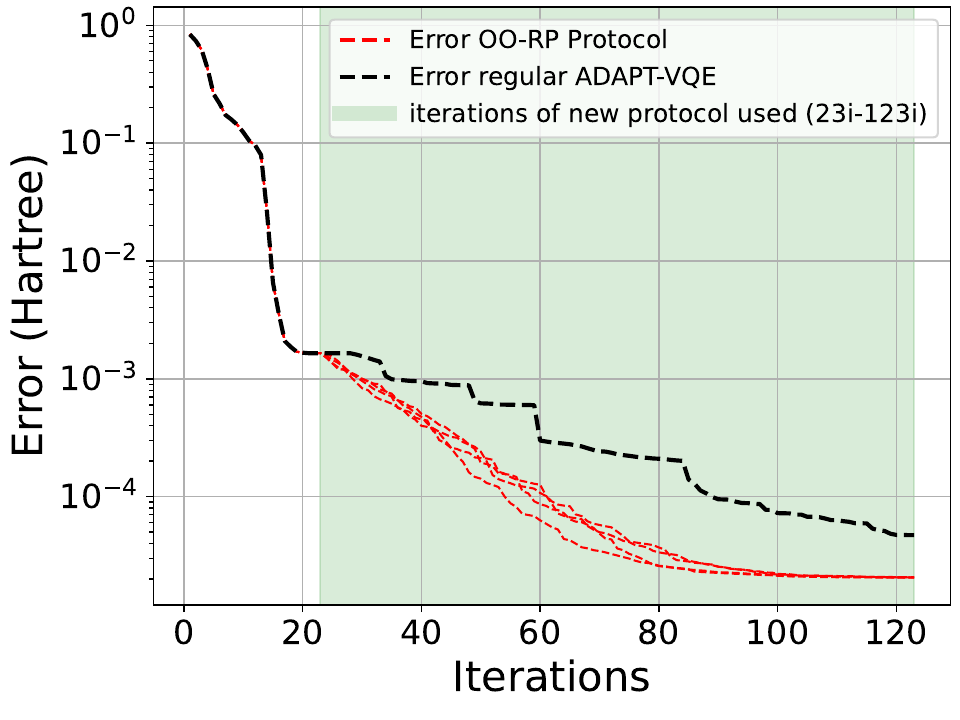}
    \caption*{\justifying (b) OO/RP Protocol (optimized operator, random position)}
\end{minipage}

\vspace{0.6cm}

\begin{minipage}{0.45\textwidth}
    \centering
    \includegraphics[width=\linewidth]{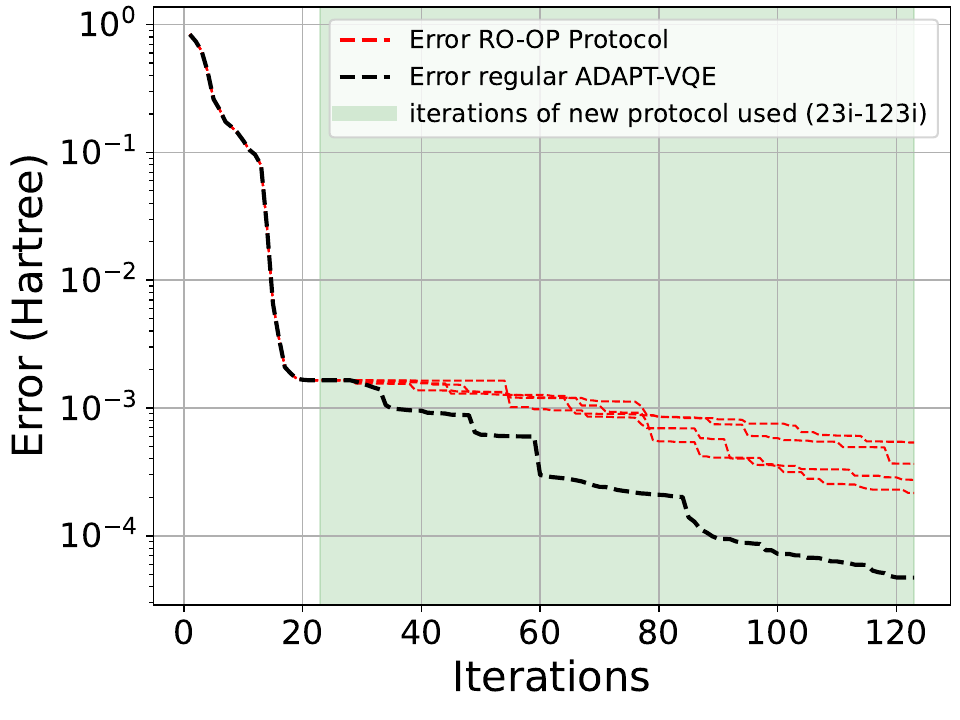}
    \caption*{\justifying (c) RO/OP Protocol (random operator, optimized position)}
\end{minipage}
\hfill
\begin{minipage}{0.45\textwidth}
    \centering
    \includegraphics[width=\linewidth]{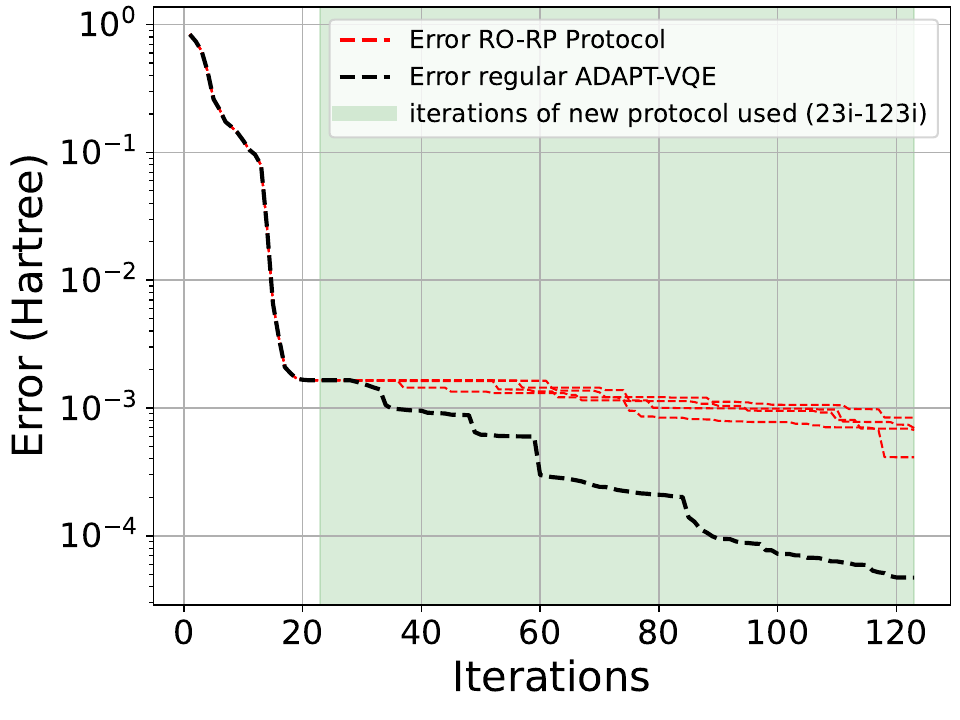}
    \caption*{\justifying (d) RO/RP Protocol (random operator, random position)}
\end{minipage}

\caption{\justifying Results of applying the four new protocols to ADAPT-VQE. The protocols start being applied three iterations after the first occurrence of a gradient trough, ensuring the algorithm is within the trough. The simulation uses the typical ADAPT-VQE (appending protocol) until iteration 22. From iteration 23 on, the enhanced protocols are activated until the algorithm converges.}
\label{fig:four_full_width_plots}
\end{figure*}

\begin{figure*}[p] 
\centering

\hspace*{0.04\textwidth}
\begin{minipage}{0.4\textwidth}
    \centering
    \includegraphics[width=\linewidth]{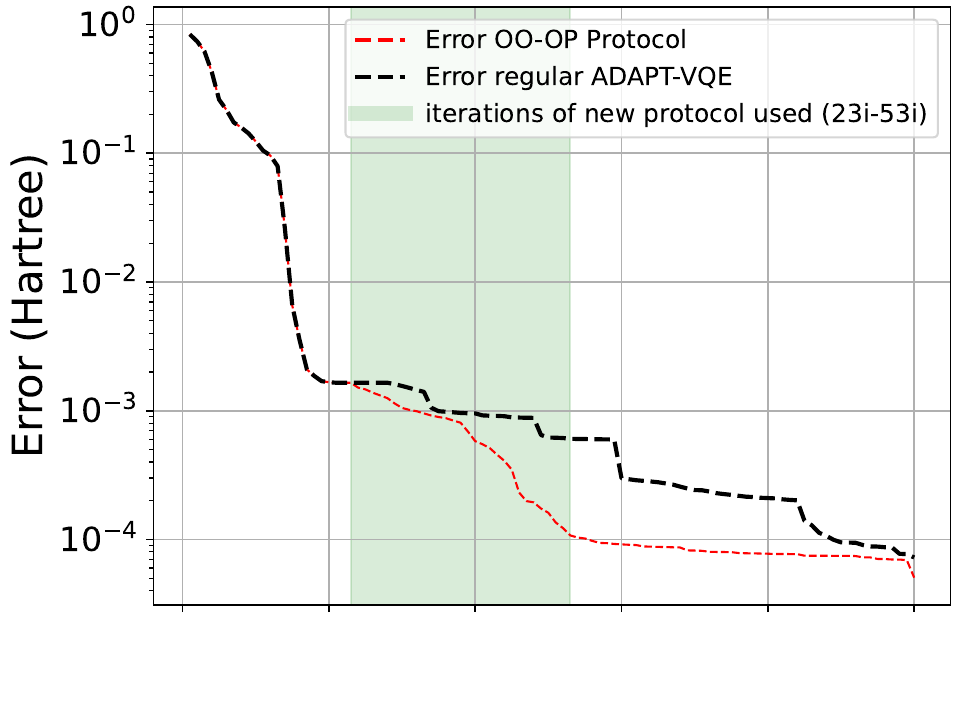}
    \phantomsubcaption
\end{minipage}
\hfill
\begin{minipage}{0.4\textwidth}
    \centering
    \includegraphics[width=\linewidth]{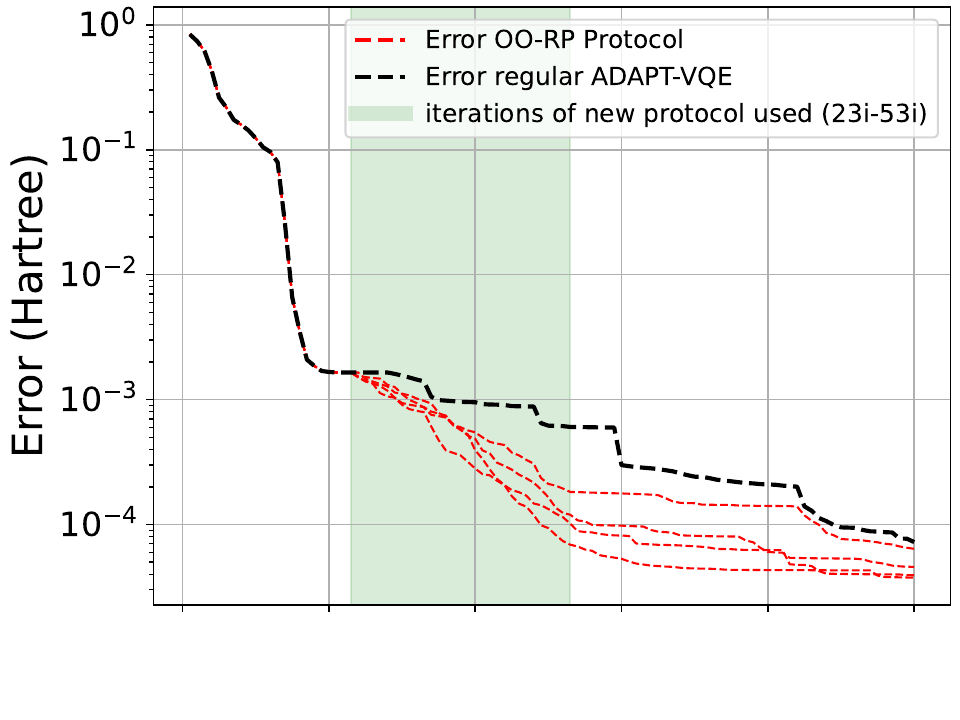}
    \phantomsubcaption
\end{minipage}
\hspace*{0.04\textwidth}

\vspace{-0.9cm}

\hspace*{0.04\textwidth}
\begin{minipage}{0.4\textwidth}
    \centering
    \includegraphics[width=\linewidth]{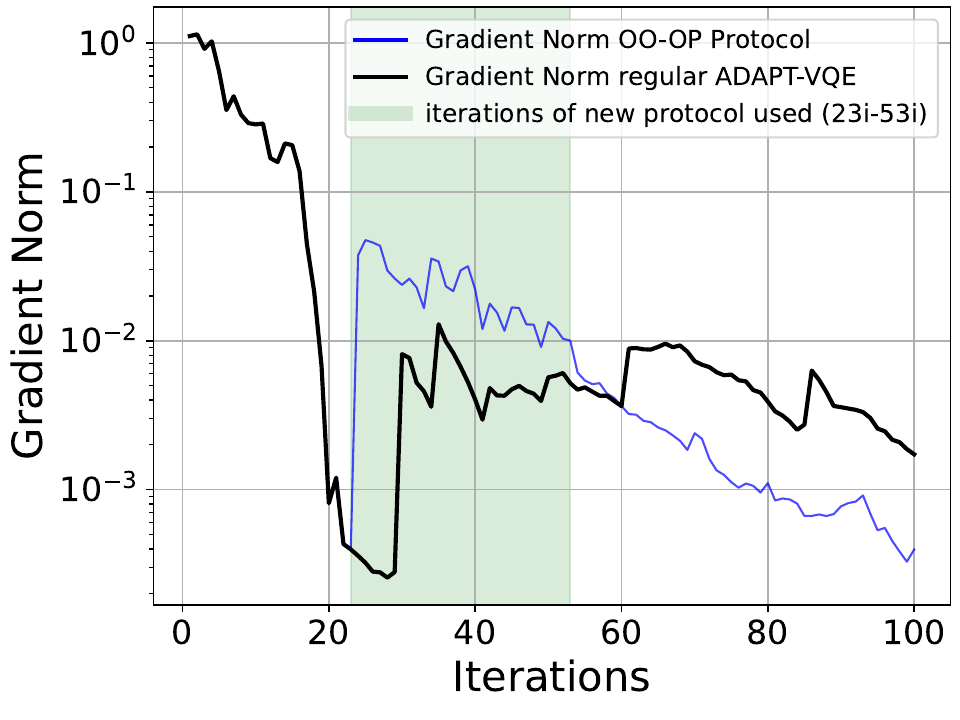}
    \phantomsubcaption
    \label{fig:protocol_5i_interv_a}
    \captionsetup{labelformat=empty , skip=-10pt}
    \caption*{\justifying (a) OO/OP Protocol (optimized operator, optimized position) used from iteration 23 to iteration 53 for a linear H$_6$ molecule with $4\text{\AA}$ interatomic distance.}
\end{minipage}
\hfill
\begin{minipage}{0.4\textwidth}
    \centering
    \includegraphics[width=\linewidth]{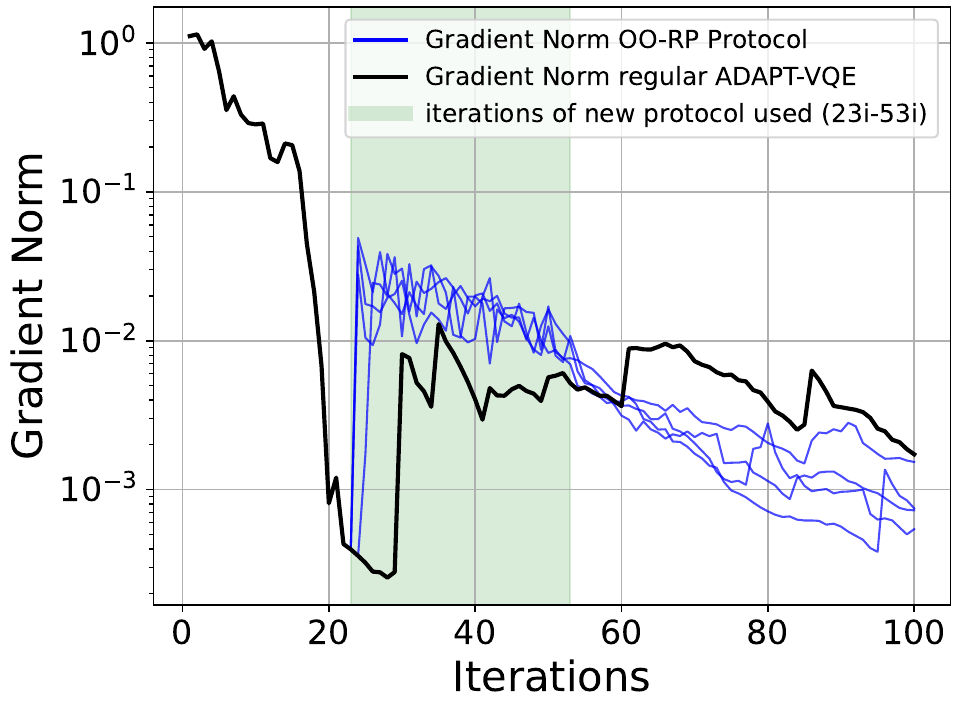}
    \phantomsubcaption
    \label{fig:protocol_5i_interv_b}
    \captionsetup{labelformat=empty , skip=-10pt}
    \caption*{\justifying (b) OO/RP Protocol (optimized operator, random position) used from iteration 23 to iteration 53.}
\end{minipage}
\hspace*{0.04\textwidth}

\vspace{0.0cm}

\hspace*{0.04\textwidth}
\begin{minipage}{0.4\textwidth}
    \centering
    \includegraphics[width=\linewidth]{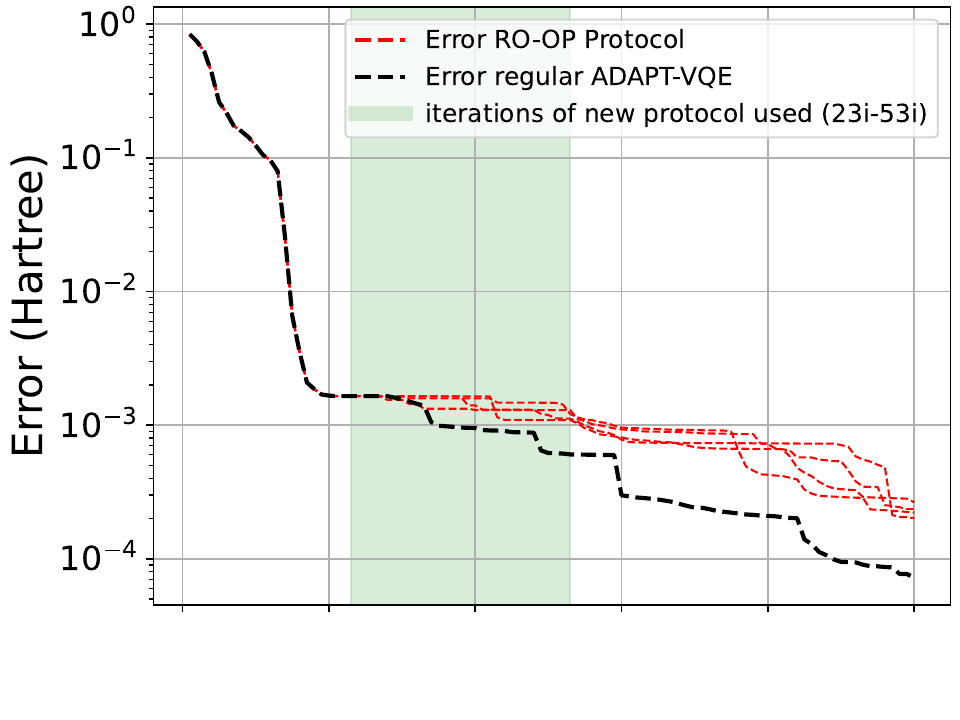}
    \phantomsubcaption
\end{minipage}
\hfill
\begin{minipage}{0.4\textwidth}
    \centering
    \includegraphics[width=\linewidth]{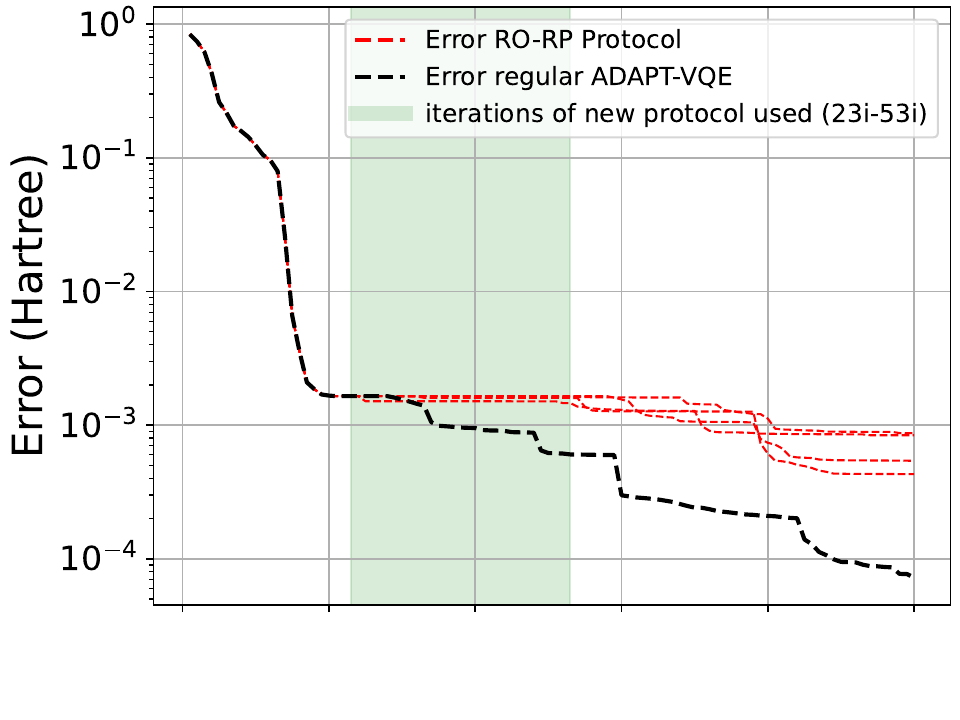}
    \phantomsubcaption
\end{minipage}
\hspace*{0.04\textwidth}

\vspace{-0.9cm}

\hspace*{0.04\textwidth}
\begin{minipage}{0.4\textwidth}
    \centering
    \includegraphics[width=\linewidth]{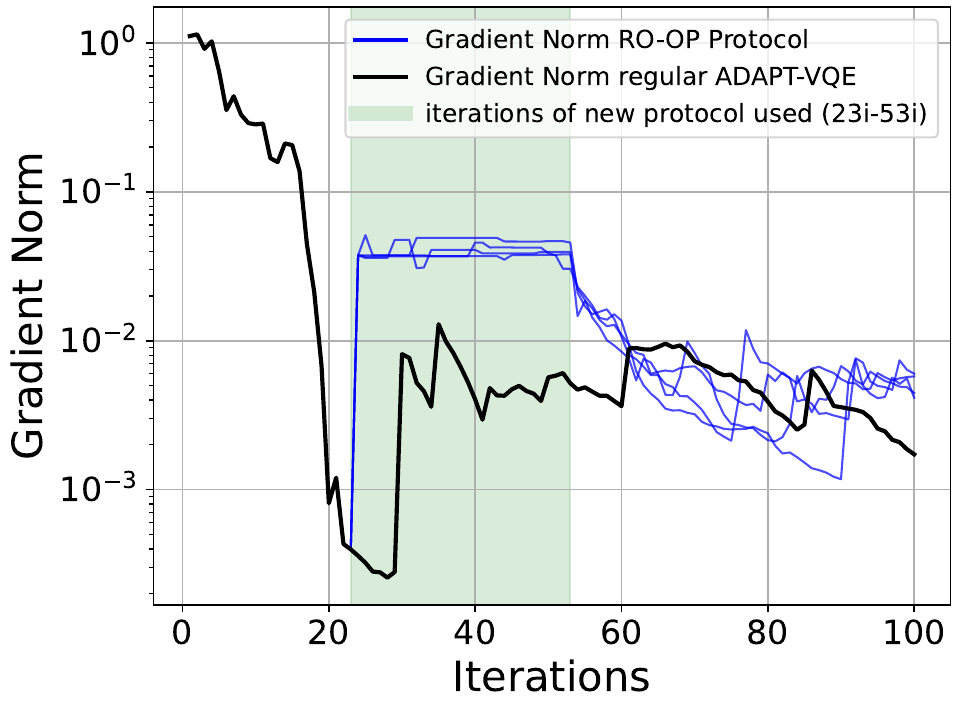}
    \phantomsubcaption
    \label{fig:protocol_5i_interv_c}
    \captionsetup{labelformat=empty , skip=-10pt}
    \caption{\justifying (c) RO/OP Protocol (random operator, optimized position) used from iteration 23 to iteration 53.}
\end{minipage}
\hfill
\begin{minipage}{0.4\textwidth}
    \centering
    \includegraphics[width=\linewidth]{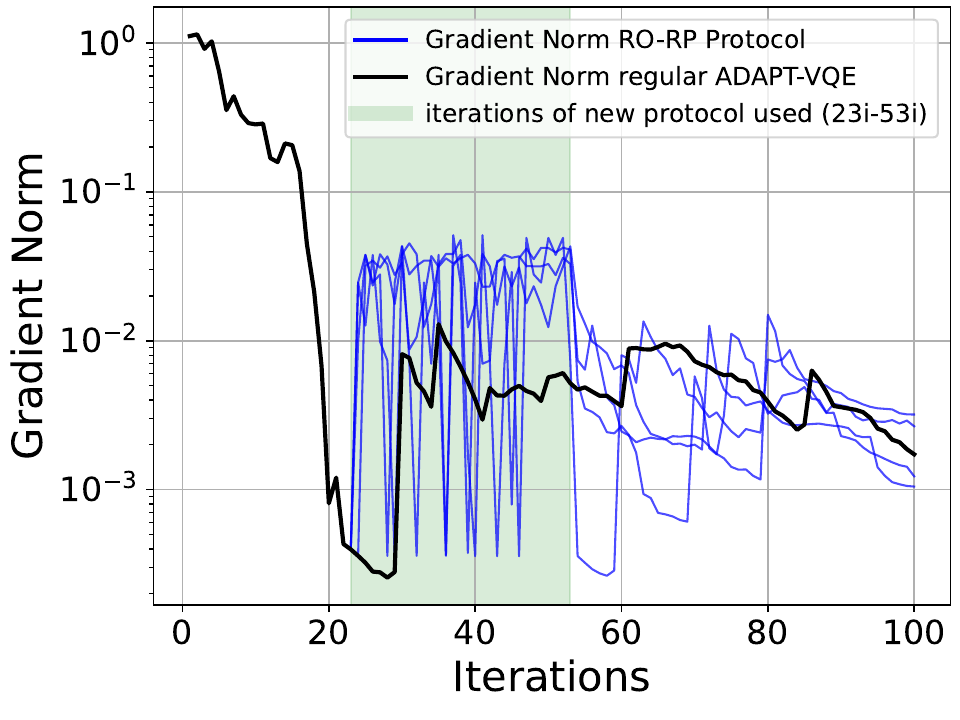}
    \phantomsubcaption
    \label{fig:protocol_5i_interv_d}
    \captionsetup{labelformat=empty , skip=-10pt}
    \caption*{\justifying (d) RO/RP Protocol (random operator, random position) used from iteration 23 to iteration 53 for a linear H$_6$ molecule with $4\text{\AA}$ interatomic distance.}
\end{minipage}
\hspace*{0.04\textwidth}

\caption{\justifying Results of applying each enhanced protocol for 30 iterations of ADAPT-VQE, from iteration 23 to iteration 53 (within the occurrence of a gradient trough). After these iterations, the algorithm continues using the normal appending protocol.}
\label{fig:five_iter_out_of_GT}
\end{figure*}

Fig.~\ref{fig:four_full_width_plots} shows the energy error as a function of the adaptive iterations for the various protocols (red lines) in comparison with standard ADAPT-VQE (black lines), where the operators are appended to the ansatz. The green regions in the plots indicate the iterations in which the new protocols have been deployed. The reason for this choice is due to the presence of a gradient trough starting at iteration $22$ for standard ADAPT-VQE, which is applied outside the green regions (causing the black and red lines to coincide prior to the green region). In practice, the detection of a gradient trough at iteration 22 can be done using the ideas discussed in Sec.~\ref{sec:Results_ID_GTs} to differentiate between gradient troughs and true convergence.

We clearly see the ability of our protocols to actively escape a gradient trough. As is evident in the figure, the OO/OP and the OO/RP protocols successfully escape gradient troughs, constantly pushing the optimization further towards the ground energy. RO/RP and RO/OP, on the other hand, are not as effective, resulting in a worse energy convergence than the canonical ADAPT-VQE.

Although the best among the new protocols seem to inherently improve energy convergence, it is conceivable to only apply them in the vicinity of a gradient trough instead of throughout the entire optimization process to avoid the additional steps required. Figure~\ref{fig:five_iter_out_of_GT} shows results for a simulation in which the new protocols are only applied for 30 iterations after the detection of a gradient trough. We plot both the energy error (top plots, red) and gradient norms (bottom plots, blue) as a function of the adaptive iterations. We can observe that the OO/OP and OO/RP schemes are able to escape troughs within the first iterations of being applied, as the norm increases rapidly with their application. After applying them, we observe a steady decrease in the gradient norm that does not present the defining features of a gradient trough. For the RO/RP and RO/OP, we observe a similar increase in the gradient norm, but the energy remains stalled. Generally, applying the proposed protocols throughout may allow us to decrease measurement costs by increasing the gradient norm (as discussed in Sec .~\ref {sec:Measurement_costs}). However, to determine if a consistent application of the protocols yields a significant advantage, a careful analysis of all costs is necessary. Such an analysis must consider all gradient measurements across all iterations, as well as all the measurements required by the optimization process. We leave this analysis for future work. 

While Fig.~\ref{fig:five_iter_out_of_GT} considers the application of our protocols for $30$ iterations, we also carried out simulations applying them for $5$ and $10$ iterations after the detection of a trough. In these cases, ADAPT-VQE generally gets trapped again after switching back to appending for all new protocols. Only when the new protocols are implemented for $30$ iterations or more is the gradient trough avoided in a sustainable manner. This is likely specific to the system we are studying here, and more investigation is required using other (molecular) systems to make a more general statement on the required iterations. That being said, the protocols we have developed in Sec.~\ref{sec:Results_ID_GTs} can always be deployed to check if the number of iterations of the new protocols is insufficient to escape the gradient trough. If it is still in the trough, it continues with the new protocol; otherwise, it reverts to the appending/regular ADAPT-VQE protocol.

We expect the OO/OP and OO/RP protocols to converge more quickly than the ADAPT-VQE protocol. This has the potential for an even greater impact on larger molecular systems, because the calculation cost required to escape a trough increases rapidly as it becomes more difficult for a strict appending procedure to affect the total ansatz when only a few operators are added and the ansatz is comparably large. On the other hand, the RO/RP and RO/OP protocols converge slower than ADAPT-VQE because they eliminate the mathematically driven selection criterion.


\section{Conclusions and Outlook}\label{sec:Conclusions}
In this work, we introduced and analyzed several modified operator selection/addition protocols for ADAPT-VQE aimed at escaping gradient troughs, a problem where the gradients with respect to all operators become simultaneously small and thus stall the ansatz improvement. Our results demonstrate that these troughs are limited to the particular ansatz position where operators are added in the original algorithm (i.e., end of the ansatz). The proposed protocols, which make use of the non-commutative properties of the exponentiated operators by introducing the possibility of adding operators in different positions where gradients are higher, significantly minimize the likelihood of getting trapped within these regions by allowing the protocol to escape the troughs. As an additional contribution, we introduce new criteria that allow ADAPT-VQE to systematically identify the occurrence of a gradient trough and distinguish it from converging, allowing one to avoid the problem of false convergence within a realistic setting. Overall, our findings show that exploring alternative positions for operator addition in ADAPT-VQE can restore its progress in situations where the original version of ADAPT-VQE stalls. This improves both robustness and practicality of ADAPT-VQE.

In addition to these advantages, our proposed protocols are able to decrease the  measurement costs of ADAPT-VQE. Even though identifying and monitoring the existence of a gradient trough requires additional gradient measurements, we design our protocols such that this overhead is negligible with respect to the costs of the original algorithm. Further, by increasing the gradient norm, our protocols reduce the accuracy required in the gradient evaluation during the operator selection step, which effectively results in a reduction in measurement costs by a factor of $10^4$ even for small (12-qubit) molecules. Additional heuristics may help further reduce costs. 

The simulations presented in this paper were carried out under idealized conditions. The performance of the algorithms may differ on real quantum computers with the presence of sampling noise, decoherence, and gate errors. We leave the study of measurement-efficient protocols and of the impact of noise for future work. Additionally, an interesting prospect for future research is developing a deeper structural understanding of when gradient troughs emerge and categorizing structural indicators for gradient troughs. This knowledge should allow us to predict when a trough emerges or enable the development of alternative ways of detecting the presence of a trough. Further, it seems promising to investigate an algorithm that combines our operator-position-based methodology with strategies to remove of irrelevant operators that have been proposed in the literature~\cite{ramoaAnsatzeNoisyVariational2022,vaquerosabater2025prunedadaptvqecompactingmolecularansatze}. While these methods are applied outside of gradient troughs, they might mitigate the gate overhead stemming from low-impact operators selected due to the stochastic criteria in the new algorithms. Finally, analyzing our protocols with different operator pools beyond the one based on qubit excitations (QE) may reveal deeper connections between pool operators, ansatz construction, and gradient geometry. This may tell us more about the fundamental workings of adaptive quantum variational algorithms.


\section{Acknowledgments}\label{sec:Thanks}

J. Stadelmann and J. \"Ubelher acknowledge support from the European Union, especially the European Union Comission, under the Erasmus+ mobility activity program. J. Stadelmann and J. \"Ubelher also thank Gerhard Mayr and Elisabeth Schmid for their assistance and commitment in securing this funding. Furthermore, gratitude is extended to the Initiative Begabung Association, which has graciously co-funded this project, in addition to its sustained commitment to fostering young students with special interests in STEM disciplines. M. Ram\^oa acknowledges support from FCT (Fundação para a Ciência e a Tecnologia), under PhD research scholarship 2022.12333.BD.  This work is in part financed by National Funds through the Portuguese funding agency, FCT, within project LA/P/0063/2020. S. E. Economou and M. Ram\^oa acknowledge support by Wellcome Leap as part of the Quantum for Bio Program. S. E. Economou and B. Sambasivam acknowledge support from the National Science Foundation (grant no. 2231328). E. Barnes acknowledges support from the Department of Energy (grant no. DE-SC0025430) and from the National Science Foundation (grant no. 2427046).

\clearpage
\bibliography{apssamp}


\appendix

\section{Generalized Gradient Formula}\label{app:AppA}
In this section, we derive the formula to obtain the gradient in any position within the ansatz.

For a typical ADAPT-VQE ansatz (Eq.~\ref{eq:ADAPT_append}), the energy expectation value is:
\begin{align}
E
&= \bra{\psi_0}
\left (\prod_{i=1}^n e^{-\theta_i A_i}\right)
\hat{H}
\left(\prod_{i=1}^n e^{\theta_{n+1-i} A_{n+1-i}}\right)
\ket{\psi_0}, \label{eq:ADAPT-energy_expectation_value}.
\end{align}
where $\ket{\psi_0}$ is the reference state. A new operator $e^{\theta_{\mu} A_{\mu}}$ can be introduced into the ansatz at any position $p$ constrained between 1 (prepending) and n+1 (appending):
\begin{equation}
\begin{aligned}
E
&= \bra{\psi_0}
\left(\prod_{i=1}^{p_s-1} e^{-\theta_i A_i}\right)
e^{-\theta_\mu A_\mu}
\hat{M} e^{\theta_\mu A_\mu} \\[1em]
&\quad
\left(\prod_{i=1}^{p_s-1} e^{\theta_{p_s-i} A_{p_s-i}}\right)
\ket{\psi_0},
\end{aligned}
\label{eq:ADAPT-energy_expectation_value_mu}
\end{equation}
where $\hat{M}$ is:

\begin{align}
\hat{M}
&:=
\left(\prod_{i=p_s}^{n} e^{-\theta_i A_i}\right)
\hat{H}
\left(\prod_{i=p_s}^{n} e^{\theta_{p_s+n-i} A_{p_s+n-i}}\right). \label{eq:M_substitute_derfi}
\end{align}
Then, the energy derivative with respect to the variational parameter $\theta_\mu$ can be written as:
\begin{equation}
\begin{aligned}
\frac{\partial E}{\partial \theta_\mu}
&=
\bra{\psi_0}
\left(\prod_{i=1}^{p_s-1} e^{-\theta_i A_i}\right)
\Bigl(-A_\mu e^{-\theta_\mu A_\mu} \,\hat{M} e^{\theta_\mu A_\mu}+\\[1em]
&\quad
e^{-\theta_\mu A_\mu} \hat{M}\,A_\mu e^{\theta_\mu A_\mu} \Bigr)
\left(\prod_{i=1}^{p_s-1} e^{\theta_{p_s-i} A_{p_s-i}}\right)
\ket{\psi_0}.
\label{eq:ADAPT-energy_expectation_value_mu_derivative}
\end{aligned}
\end{equation}
Setting $\theta_\mu = 0$ (typical ADAPT-VQE initialization) and using the commutator $[A_\mu, \hat{M}]=A_\mu\hat{M}-\hat{M}A_\mu$ this can be simplified to
\begin{align}
\left.\frac{\partial E}{\partial \theta_\mu}\right|_{\theta_\mu=0}
&=
\bra{\psi_0}
\left(\prod_{i=1}^{p_s-1} e^{-\theta_i A_i}\right)
\bigl[\hat{M},A_\mu\bigr]\nonumber\\
&\qquad
\left(\prod_{i=1}^{p_s-1} e^{\theta_{p_s-i} A_{p_s-i}}\right)
\ket{\psi_0}.
\label{eq:ADAPT-energy_expectation_value_simplified}
\end{align}
By expanding each ansatz side we obtain:

\begin{align}
\left.\frac{\partial E}{\partial \theta_\mu}\right|_{\theta_\mu=0}
&=
\bra{\psi_n}
\Bigg[\hat{H},\;
\left(\prod_{i=p_s}^{n} e^{\theta_{p_s+n-i} A_{p_s+n-i}}\right)
A_\mu \nonumber\\[1em]
&\quad
\left(\prod_{i=p_s}^{n} e^{-\theta_i A_i}\right)
\Biggr]
\ket{\psi_n}
\label{eq:ADAPT-energy_expectation_value_final}
\end{align}
with $\ket{\psi}$ being defined as the wavefunction of the ansatz prior to the insertion of the new operator. In this notation, this parametrized wavefunction would be defined as:

\begin{equation}
\begin{aligned}
\ket{\psi_n}
&=
\left(\prod_{i=1}^{n} e^{\theta_{1+n-i} A_{1+n-i}}\right) \ket{\psi_0}.
\end{aligned}
\label{eq:ADAPT_ansatz_before_inserting_operator}
\end{equation}

In order to implement this gradient measurement circuit for arbitrary positions other than appending in a quantum computer, the parameter-shift rule~\cite{Mitarai_2018} needs to be applied on Eq.~\ref{eq:ADAPT-energy_expectation_value_simplified}. In the context of qubit or fermionic excitations operating on real states, as is the case in this study, it has been demonstrated that the expense associated with measuring each gradient is approximately equal to the cost of measuring the energy. This is accomplished by implementing a constant amount of supplementary gates, a phenomenon comparable to that observed in the context of a single Pauli string~\cite{D0SC06627C}. This results in total cost in the order of $\mathcal{O}(N^8)$ for measuring the gradient in any arbitrary ansatz position.


\section{Pseudo Codes}\label{app:AppB}

In this section, we include pseudo-code for all protocols proposed in Sec.~\ref{sec:Results_new_protocols}. We start with the standard ADAPT-VQE protocol (Algorithm~\ref{alg:appending_adapt_vqe}) as a baseline, and follow with the new protocols: optimized operator optimized position (OO/OP, Algorithm~\ref{alg:oo-op_adapt_vqe}), optimized operator random position (OO/RP, Algorithm~\ref{alg:oo-rp_adapt_vqe}), random operator optimized position (RO/OP Algorithm~\ref{alg:ro-op_adapt_vqe}) and random operator random position (RO/RP Algorithm~\ref{alg:ro-rp_adapt_vqe}). Furthermore, the subroutines max\_gradients\_appending (Alg.~\ref{alg:max_gradients}) and VQE (Alg.~\ref{alg:VQE_subroutine}) are included. The function optimize within the VQE subroutine makes use of classical hardware to propose optimization parameters, while the expectation values with the adjusted parameters are then evaluated on quantum hardware with the aim to minimize the energy expectation value. This is repeated until either the optimizer converges or the maximum number of the optimizer iterations is reached.

\begin{algorithm}[H]
\caption{max\_gradients\_appending($n_\text{ops}$, $\ket{\psi}$,
$\hat{H}$, $\mathbf{A}$)}
\label{alg:max_gradients}
\KwIn{$n_\text{ops}$: number of operators to output; $\ket{\psi}$: current ansatz state; $\hat{H}$: Hamiltonian; $\mathbf{A}$: operator pool}
\KwOut{$n_\text{ops}$ operator indices with highest gradient magnitudes;\ corresponding gradients}

\ForEach{$A_i \in \mathbf{A}$}{
        $g_i \leftarrow \bra{\psi} [\hat{H}, A_{i}] \ket{\psi}$\;
    }
$\text{max}_{\text{gradients}} \gets \{g_i\} \text{ sorted in descending order of } |g_i|$

max$_\text{indices} \gets$ Elements of $\mathbf{A}$ sorted according to max$_\text{gradients}$;

\Return{max$_\text{gradients}[:n_\text{ops}]$, max$_\text{indices}[:n_\text{ops}]$}
\end{algorithm}

\begin{algorithm}[H]
\caption{VQE($\hat{H}$, $\ket{\psi_{j+1}}$, $\ket{\psi_j}$, $p$)}
\label{alg:VQE_subroutine}
\KwIn{$\hat{H}$: Hamiltonian; $\ket{\psi_{j+1}}$: current ansatz state; $\ket{\psi_{j}}$: previous ansatz state; $p$: position of new operator in the ansatz}
\KwOut{ansatz state with optimized parameters; optimized variational energy}
$\vec{\theta} \leftarrow \text{from } \ket{\psi_j}$\;
$\vec{\theta} \leftarrow (\vec{\theta}_{1:p},\,0,\,\vec{\theta}_{p+1:n})$\;
$\ket{\psi_{j+1}(\vec{\theta})} \leftarrow \ket{\psi_{j+1}},\vec{\theta}$\;
$\ket{\psi_{j+1}(\vec{\theta})} \leftarrow \text{optimizer} \big( \; \bra{\psi_{j+1}(\vec{\theta})} \hat{H} \ket{\psi_{j+1}(\vec{\theta})} \; \big)$\;
$E(\vec{\theta}) = \bra{\psi_{j+1}(\vec{\theta})} \hat{H} \ket{\psi_{j+1}(\vec{\theta})}$\;

\Return{$\ket{\psi_{j+1}(\vec{\theta})}, E(\vec{\theta})$}
\end{algorithm}

\begin{algorithm}[H]
\caption{ADAPT-VQE Appending Algorithm}
\label{alg:appending_adapt_vqe}

$j \leftarrow 0$\;
$\mathbf{A} \leftarrow \text{Operator pool}$\;
$\ket{\psi_0} \leftarrow \text{Hartree-Fock reference state}$\;

\While{not Converged}{
    $\mu, g \leftarrow \texttt{max\_gradients\_appending}(1, \ket{\psi_j}, \hat{H}, \mathbf{A})$\;
    $\ket{\psi_{j+1}} \leftarrow e^{\theta_\mu A_\mu} \ket{\psi_j}$\;
    $\ket{\psi_{j+1}},\, E_{j+1} \leftarrow \texttt{VQE} \bigl(\hat{H},\, \ket{\psi_{j+1}},\, \ket{\psi_j},\, n+1\bigr)$
    $j \leftarrow j + 1$\;
}

\KwOut{$\ket{\psi_{j+1}},\, E_{j+1}$}

\end{algorithm}

\begin{algorithm}[H]
\caption{Optimized Operator / Optimized Position (OO/OP) ADAPT-VQE}
\label{alg:oo-op_adapt_vqe}

$j \leftarrow 0$\;
$\mathbf{A} \leftarrow \text{Operator pool}$\;
$\ket{\psi_0} \leftarrow \text{Hartree-Fock reference state}$\;

\While{not Converged}{
    $\mu^{[10]}, g^{[10]} \leftarrow \texttt{max\_gradients\_appending}(10, \ket{\psi_j}, \hat{H}, \mathbf{A})$\;
    $\vec{\theta} \leftarrow \text{from } \ket{\psi_j}$\;

    \ForEach{$i \text{ in } 1:10$}{
        \For{$p = 1, \dots, n+1$}{
            $U_{p:n} \leftarrow \left(\prod_{i=p}^{n} e^{\theta_{p+n-i} A_{p+n-i}}\right)$\;
            $g_{i,p} \leftarrow \bra{\psi_j} \Bigg[\hat{H}, U_{p:n} A_{\mu_i} U_{p:n}^\dagger \Biggr] \ket{\psi_j}$\;
        }
        $g_i^\text{max}, p^\text{opt} \leftarrow \max_{p} |g_{i,p}|$\;
    }
    
    $\mu \leftarrow \arg\max_i \{ g_i^\text{max} \}$\;
    $p_s \leftarrow p^\text{opt}$\;

    $\begin{aligned}
    \ket{\psi_{j+1}} \leftarrow\;&
    \Bigg[\left(\prod_{i=p_s}^{n} e^{\theta_{p_s+n-i} A_{p_s+n-i}}\right)
    e^{\theta_\mu A_\mu} \\
    &
    \left(\prod_{i=1}^{p_s-1} e^{\theta_{p_s-i} A_{p_s-i}}\right)
    \ket{\psi_0}\Bigg];
    \end{aligned}$
    
    $\ket{\psi_{j+1}},\, E_{j+1} \leftarrow \texttt{VQE} \bigl(\hat{H},\, \ket{\psi_{j+1}},\, \ket{\psi_j},\, p_s)$
    
    $j \leftarrow j + 1$\;
}

\KwOut{$\ket{\psi_{j+1}},\, E_{j+1}$}

\end{algorithm}

\begin{algorithm}[H]
\caption{Optimized Operator / Random Position (OO/RP) ADAPT-VQE}
\label{alg:oo-rp_adapt_vqe}
$j \leftarrow 0$\;
$\mathbf{A} \leftarrow \text{Operator pool}$\;
$\ket{\psi_0} \leftarrow \text{Hartree-Fock reference state}$\;

\While{not Converged}{
    $p_s \gets \text{random element from } \{1, \dots, n+1\}$\;
    $\vec{\theta} \leftarrow \text{from } \ket{\psi_j}$\;
    \ForEach{$A_i \in \mathbf{A}$}{
        $U_{p:n} \leftarrow \left(\prod_{i=p}^{n} e^{\theta_{p+n-i} A_{p+n-i}}\right)$\;
        $g_{i,p_s} \leftarrow \bra{\psi_j} \Bigg[\hat{H}, U_{p_s:n} A_i U_{p_s:n}^\dagger \Biggr] \ket{\psi_j}$\;
    }
    $\mu \leftarrow \arg\max_i \{ |g_{i,p_s}| \}$\;
    
    $\begin{aligned}
    \ket{\psi_{j+1}} \leftarrow\;&
    \Bigg[\left(\prod_{i=p_s}^{n} e^{\theta_{p_s+n-i} A_{p_s+n-i}}\right)
    e^{\theta_\mu A_\mu} \\
    &
    \left(\prod_{i=1}^{p_s-1} e^{\theta_{p_s-i} A_{p_s-i}}\right)
    \ket{\psi_0}\Bigg];
    \end{aligned}$
    
    $\ket{\psi_{j+1}},\, E_{j+1} \leftarrow \texttt{VQE} \bigl(\hat{H},\, \ket{\psi_{j+1}},\, \ket{\psi_j},\, p_s)$
    
    $j \leftarrow j + 1$\;
}

\KwOut{$\ket{\psi_{j+1}},\, E_{j+1}$}
\end{algorithm}

\begin{algorithm}[H]
\caption{Random Operator / Optimized Position (RO/OP) ADAPT-VQE}
\label{alg:ro-op_adapt_vqe}

$j \leftarrow 0$\;
$\mathbf{A} \leftarrow \text{Operator pool}$\;
$\ket{\psi_0} \leftarrow \text{Hartree-Fock reference state}$\;

\While{not Converged}{
    $\mu^{[10]}, g^{[10]} \leftarrow \texttt{max\_gradients\_appending}(10, \ket{\psi_j}, \hat{H}, \mathbf{A})$\;
    $\vec{\theta} \leftarrow \text{from } \ket{\psi_j}$\;

    \ForEach{$i\in 1:10$}{
        \For{$p = 1, \dots, n+1$}{
            $U_{p:n} \leftarrow \left(\prod_{i=p}^{n} e^{\theta_{p+n-i} A_{p+n-i}}\right)$\;
            $g_{i,p} \leftarrow \bra{\psi_j} \Bigg[\hat{H}, U_{p:n} A_{\mu_i} U_{p:n}^\dagger \Biggr] \ket{\psi_j}$\;
        }
        $g_i^\text{max}, (p)_i^\text{opt} \leftarrow \max_{p} |g_{i,p}|$\;
    }
    $g^{\text{max}},i_0 \leftarrow \max_ig_i^\text{max}$\;
    $p_s\leftarrow (p)_{i_0}^{\text{opt}}$\;
    $A_{\mu} \leftarrow \text{random element from $\mathbf{A}$}$\;
    $\begin{aligned}
    \ket{\psi_{j+1}} \leftarrow\;&
    \Bigg[\left(\prod_{i=p_s}^{n} e^{\theta_{p_s+n-i} A_{p_s+n-i}}\right)
    e^{\theta_\mu A_\mu} \\
    &
    \left(\prod_{i=1}^{p_s-1} e^{\theta_{p_s-i} A_{p_s-i}}\right)
    \ket{\psi_0}\Bigg];
    \end{aligned}$
    
    $\ket{\psi_{j+1}},\, E_{j+1} \leftarrow \texttt{VQE} \bigl(\hat{H},\, \ket{\psi_{j+1}},\, \ket{\psi_j},\, p_s)$
    
    $j \leftarrow j + 1$\;
}

\KwOut{$\ket{\psi_{j+1}},\, E_{j+1}$}

\end{algorithm}

\begin{algorithm}[H]
\caption{Random Operator / Random Position (RO/RP) ADAPT-VQE}
\label{alg:ro-rp_adapt_vqe}
\SetAlgoLined
$j \gets 0$\;
$\mathbf{A} \leftarrow \text{Operator pool}$\;
$\ket{\psi_0} \leftarrow \text{Hartree-Fock reference state}$\;

\While{not Converged}{
    $\vec{\theta} \leftarrow \text{from } \ket{\psi_j}$\;
    $A_\mu \gets \text{random element from } A$\;
    $p_s \gets \text{random element from } \{1, \dots, n+1\}$\;
    
    $\begin{aligned}
    \ket{\psi_{j+1}} \leftarrow\;&
    \Bigg[\left(\prod_{i=p_s}^{n} e^{\theta_{p_s+n-i} A_{p_s+n-i}}\right)
    e^{\theta_\mu A_\mu} \\
    &
    \left(\prod_{i=1}^{p_s-1} e^{\theta_{p_s-i} A_{p_s-i}}\right)
    \ket{\psi_0}\Bigg];
    \end{aligned}$
    
    $\ket{\psi_{j+1}},\, E_{j+1} \leftarrow \texttt{VQE} \bigl(\hat{H},\, \ket{\psi_{j+1}},\, \ket{\psi_j},\, p_s)$
    
    $j \gets j + 1$\;
}
\KwOut{$\ket{\psi_{j+1}},\, E_{j+1}$}
\end{algorithm}


\section{Relation Between Gradient Troughs and Gradients Across the Ansatz}\label{app:AppC}

\begin{figure}[!htbp]
    \centering
    \begin{subfigure}[b]{0.45\textwidth}
        \centering
        \includegraphics[width=\textwidth]{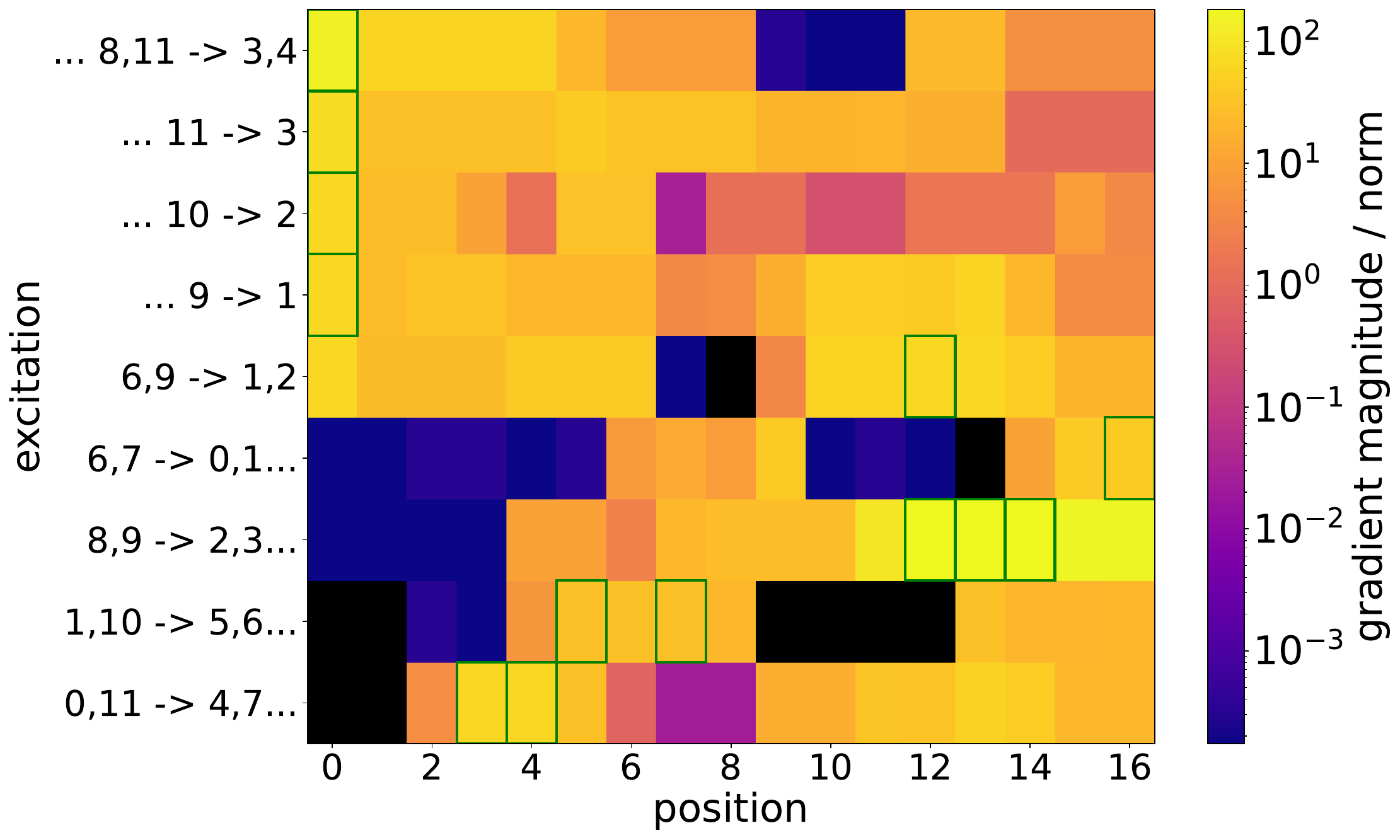}
        \caption{\justifying iteration 16 - ahead of gradient trough}
        \label{fig:grad_i16}
    \end{subfigure}
    
    \vspace{0.5em} 

    \begin{subfigure}[b]{0.45\textwidth}
        \centering
        \includegraphics[width=\textwidth]{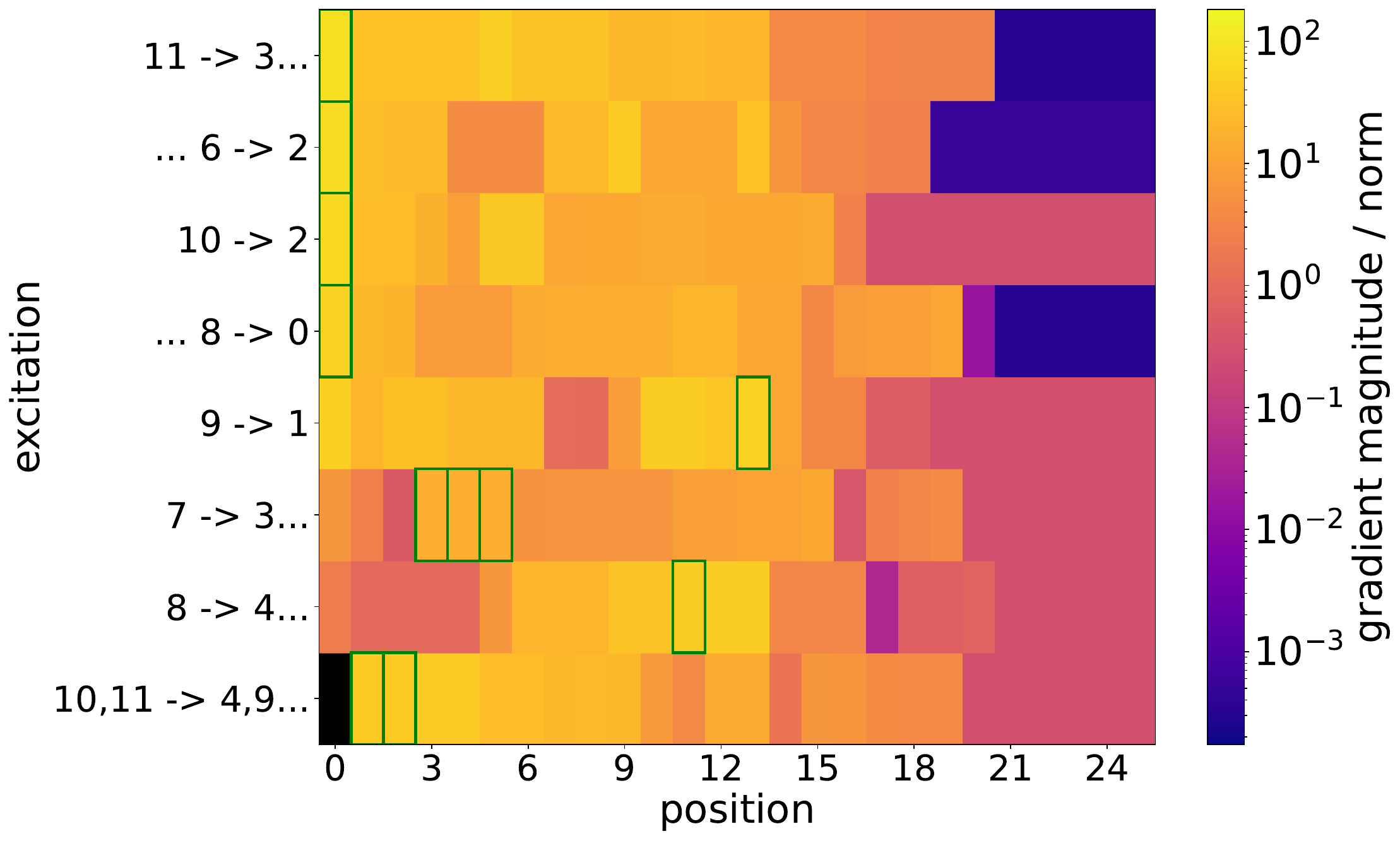}
        \caption{\justifying iteration 25 - within the gradient trough}
        \label{fig:grad_i25}
    \end{subfigure}
    
    \vspace{0.5em} 
    
    \begin{subfigure}[b]{0.45\textwidth}
        \centering
        \includegraphics[width=\textwidth]{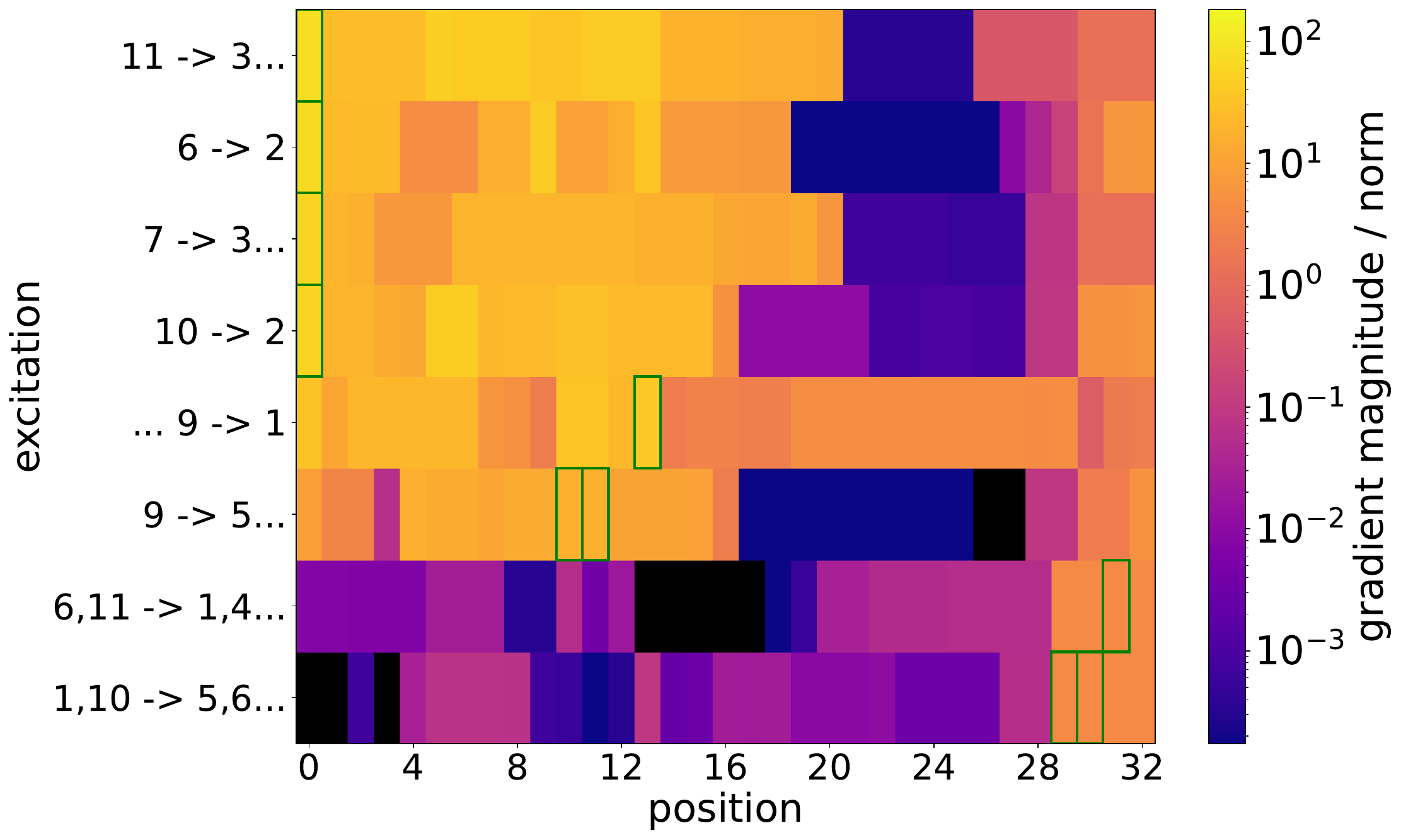}
        \caption{\justifying iteration 32 - after the gradient trough}
        \label{fig:grad_i32}
    \end{subfigure}

    \caption{\justifying We compare the normalized gradient magnitudes before (iteration 16), during (iteration 25), and after (iteration 32) a gradient trough for the specified operators at all ansatz positions during a regular ADAPT-VQE execution (appending) for the top 5 operators for the prepending and appending positions. Top operators for prepending are labeled with `...[source orbitals] $\rightarrow$ [target orbitals]' and the ones for appending with `[source orbitals] $\rightarrow$ [target orbitals] ...'. Operators denoted as `[source orbitals] $\rightarrow$ [target orbitals]' are within the top for appending and prepending. The green boxes highlight the highest gradient magnitude position for the operator investigated.}
    \label{fig:grid_plots}
\end{figure}

\begin{figure}[ht]
    \centering
    \includegraphics[width=0.45\textwidth]{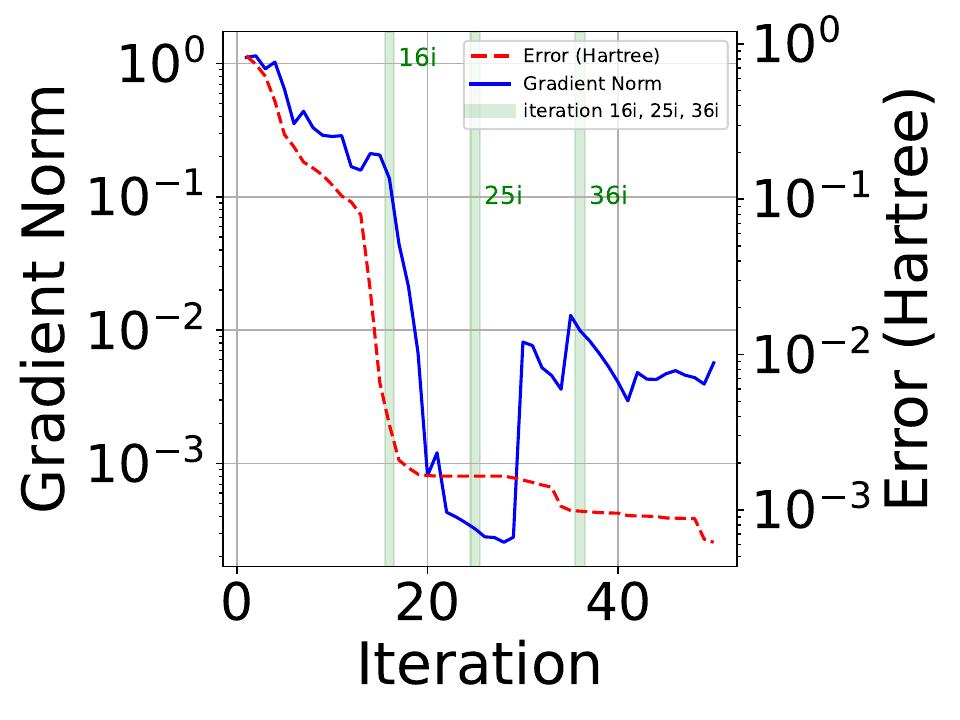}
    \caption{\justifying Iterations considered in Fig.~\ref{fig:grid_plots} are highlighted. The gradient trough occurs in a regular ADAPT-VQE execution (appending operators).}
    \label{fig:marked_iterations_gt_for_grid_plots}
\end{figure}

In this appendix, we further analyze the behavior of the gradients over the positions as discussed in Sec.~\ref{sec:Results_ID_GTs}. Figure~\ref{fig:grid_plots} shows a comparison of normalized gradient magnitudes before (iteration 16), within (iteration 25) and after (iteration 32) a gradient trough for the specified operators over all ansatz positions within a regular ADAPT-VQE execution (operators are always appended to the end of the ansatz). The gradients magnitudes are normalized by the minimal gradient magnitude across positions for iteration 28. We consider the 5 operators with largest gradients when appending and when prepending. When prepending, we write `...[source orbitals] $\rightarrow$ [target orbitals]'; when appending, we write `[source orbitals] $\rightarrow$ [target orbitals] ...'. Operators represented by `[source orbitals] $\rightarrow$ [target orbitals]' ranked among the top 5 both for prepending and for appending. The green boxes highlight the highest gradient magnitude position for the operator under scrutiny. The specific iterations which are represented within Fig.\ref{fig:grid_plots} are highlighted in  Fig.\ref{fig:marked_iterations_gt_for_grid_plots}, which showcases the conventional appending ADAPT-VQE calculation for the initial 50 iterations, accompanied by a gradient trough spanning from iteration 20 to iteration 29.

At iteration 16 (Fig.~\ref{fig:grad_i16}), before the gradient trough, the operators with the highest gradients for prepending/appending have near-maximum gradients at that same position. In what concerns the top 5 operators for appending, the gradients typically vanish when we consider prepending them (or more generally, placing them very early on the ansatz). This behavior is expected: when we select an operator due to it having a high gradient when appending/prepending, this will tend to be the position where it performs the best. However, this is not the case during a gradient trough, as we can observe in Fig.~\ref{fig:grad_i25}. In this case, even the top 5 operators selected for appending have significantly higher gradients when they are prepended. This can be explained by the commutator argumentation in Sec.~\ref{sec:Results_ID_GTs}. Finally, after leaving the gradient trough at iteration 32 (Fig.~\ref{fig:grad_i32}), the trend reverts back to what we observed before the gradient trough was encountered: The operators with highest gradients for appending have near-maximum gradients in the appending position.

\clearpage
\end{document}